\documentclass[11pt]{article}
\usepackage{amsmath, amssymb}
\usepackage{slashed}
\usepackage{verbatim}
\usepackage{ulem} 
\usepackage{latexsym}
\usepackage{euscript}
\usepackage{amsfonts}
\usepackage{verbatim}
\usepackage[]{array}
\usepackage[]{mathrsfs}
\usepackage[]{tensor}
\usepackage{graphicx}
\usepackage{amsmath}
\usepackage{amssymb}
\usepackage{amsxtra}
\usepackage{color}
\def\be{\begin{eqnarray}}
\def\ee{\end{eqnarray}}
\def\0{\nonumber}

\def\tr{{\rm tr}} \def\Tr{{\rm Tr}}

\def\det{\rm det}
\usepackage{slashed}
\usepackage{verbatim}
\usepackage{latexsym}\usepackage{color}
\usepackage{euscript}

\newcommand\EV{\EuScript{V}}
\newcommand\EF{\EuScript{F}}

\normalfont\large

\begin{document}
\flushright{SISSA  17/2021/FISI}
\flushleft
\vskip 2cm
\begin{center}

{\LARGE Perturbative and non-pertrubative trace anomalies }

\vskip 1cm

{\large  L.~Bonora
\\
\textit{ International School for Advanced Studies (SISSA),\\Via
Bonomea 265, 34136 Trieste, Italy   }
 }
\end{center}
\vskip 1cm
{\bf Abstract}. We study the definition of trace anomalies for models of Dirac and Weyl fermions 
coupled to a metric and a gauge potential. While in the non-perturbative case the trace anomaly is the response of the effective action to a Weyl transformation, the definition in a perturbative approach is more involved. In the latter case we use a specific formula proposed by M.Duff, of which we present a physical interpretation. The main body of the paper consists in deriving trace anomalies with the above formula and comparing them with the corresponding non-perturbative results. We show that they coincide and stress the basic role of diffeomorphism invariance for the validity of the approach.   

\vskip 1cm

\section{Introduction. A perturbative definition of trace anomaly.}

 Chiral anomalies appear in the divergence of chiral currents as a result of regularizing fermion loops, in the perturbative case, and effective actions, in non-perturbative approaches. No essential ambiguities arise in the process of defining the relevant procedures.  The case of trace anomalies is somewhat more involved (for various derivations of trace anomalies see \cite{capper}-\cite{parker}). As we shall see on examples, {{\it a priori} several definitions of the trace anomaly are  possible in a perturbative approach, definitions that lead to different results}.  As we shall see, there are in  fact several sources of ambiguity. {To eliminate these ambiguities,} our definition of trace anomaly in the perturbative case will be the one adopted by M.Duff, see for instance \cite{duffrecent}. If $T_{\mu\nu}$ is the stress-energy tensor of  a theory, the trace anomaly is given by the difference
\be
{ g^{\mu\nu}\langle\!\langle T_{\mu\nu}(x) \rangle\!\rangle -\langle\!\langle g^{\mu\nu} T_{\mu\nu}(x) \rangle\!\rangle}\label{Duff}
\ee
This formula is not by itself self-explanatory and needs suitable specifications, which will be given in due time below. But, just to make an initial example, when a theory is conformal invariant, the field operator $T_\mu^\mu(x)$ vanishes on shell, while in the case a theory contains a conformal soft breaking term (a mass term, for instance) $T_\mu^\mu(x)\neq 0$ even on-shell. The second term of \eqref{Duff} is {certainly present in such a case and the subtraction} is needed in order to exclude this unwanted term from the anomaly. As a matter of fact, as we shall see, this term is non-vanishing in many other instances, { and in a subtler way (for more down-to-earth uncertainties, see \cite{BS} and references therein) }. The purpose of our paper is to discuss the application of formula \eqref{Duff}. To study this issue we have to enlarge our vantage point by considering not only odd parity trace anomalies, but also even parity ones. We will focus, in particular, on the anomalies of a free fermion field model coupled to an Abelian  vector potential $V_\mu$ with curvature $F_{\mu\nu}$. Beside the odd parity anomaly with density
$\epsilon^{\mu\nu\lambda\rho}F_{\mu\nu} F_{\lambda \rho}$ we wish to consider also the even parity anomaly with density $F_{\mu\nu} F^{\mu\nu}$.  These density can show up as trace anomalies in the form $\int d^4x \sqrt{g}\, \omega(x)\,  F_{\mu\nu}(x) F^{\mu\nu}(x)$ and $\int d^4x \sqrt{g}\, \omega(x)\,\epsilon^{\mu\nu\lambda\rho}  F_{\mu\nu}(x) F_{\lambda\rho}(x)$, since they are consistent under the conformal transformations $\delta_\omega g_{\mu\nu} =2\omega g_{\mu\nu}$ and $\delta_\omega V_\mu=0$. We will consider two examples where these anomalies do appear: the even trace anomaly in the theory of a Dirac fermion, and the odd one in the theory of a Weyl fermion, both coupled to a vector potential $V_\mu$. {In both cases we compare the results with the ones obtained via the non-perturbative heat-kernel-like method, which we refer to as the Seeley-DeWitt (SDW) method, in which case the trace anomaly definition is the standard one, i.e. the response of the effective action under a Weyl transformation. In both cases the results coincide.} 

{ A pertinent question is why in the perturbative approach there is room for ambiguities. The reason is that the approach based on Feynman diagrams is reasonably viable at the lowest order of the perturbative expansion. Were we able to compute  higher order approximations there would be no room for ambiguities. Unfortunately higher order calculations are much less accessible, in general, and we have to make do with the lowest order. Now, at the lowest order, the relevant cohomology is determined by the lowest order of the perturbative cohomology (see Appendix B). The lowest order cohomology is {\it too simple}, for instance it contains many more cocycles than the non-perturbative (complete) cohomology. The definition \eqref{Duff} is more {\it balanced} than each of the two terms separately, it forces the perturbative calculations in the right non-perturbative direction. It offers also a clear physical interpretation, which we discuss in the conclusive section.}

The paper is organized as follows. In section 2 we derive the above-mentioned even trace anomaly for a Dirac fermion coupled to vector potential with a perturbative method (Feynman diagrams plus dimensional regularization). In section 3 we derive the same result via the Seeley-DeWitt method. Section 4 is devoted to the odd parity trace anomaly for a Weyl fermion coupled to a vector potential and section 5 to the same derivation via SDW. In section 2.5 and in section 6 we discuss the conservation of diffeomorphims in presence of a gauge vector field in the two cases, respectively. In section 7 we make some final comments and, in particular, we offer our interpretation of the definition \eqref{Duff} for the trace anomaly.
In Appendix A we show that the density $F_{\mu\nu}F^{\mu\nu}$ can never appear as an anomaly of the divergence of a vector or axial current. In Appendix B we define perturbative cohomology.

 \section{Even parity trace anomalies due to vector gauge field}

We consider the action of a Dirac fermion coupled to a metric and an Abelian vector potential
\be
S= \int d^4x \, \sqrt{g} \, i\overline {\psi} \gamma^\mu\left(\partial_\mu +\frac
12
\omega_\mu+ V_\mu \right)\psi \label{actionS}
\ee
where $\gamma^\mu= e_a^\mu\, \gamma^a$, and  $\omega_\mu$ is the spin connection, $\omega_\mu= \omega_\mu^{ab} \Sigma_{ab}$ and
$\Sigma_{ab} = \frac 14 [\gamma_a,\gamma_b]$ are the Lorentz generators.
For a uniform treatment with the non-Abelian case where the generators are anti-Hermitean, we use an imaginary vector field $V_\mu$. Eventually one can make the replacement $V_\mu \to -i V_\mu$.
The vector current is $j_\mu= i\overline \psi \gamma_\mu \psi$ and the stress-energy tensor is 
\be
T_{\mu\nu}= \frac i4 \overline {\psi} \gamma_\mu {\stackrel{\leftrightarrow}{\nabla}}_\nu\psi+ \{\mu\leftrightarrow \nu\}
 \label{emt}
\ee
They are both conserved on shell. $T_{\mu\nu}$ is also traceless on shell.
In this section we use perturbative methods with dimensional regularization and focus on the possibility for $F_{\mu\nu} F^{\mu\nu}$ to appear as a trace anomaly. {As expected such a density cannot appear in the divergence of a current, as we show anyhow, in Appendix A.}

Before we start a clarification is in order concerning the definition \eqref{emt} of the em tensor. There is in fact another definition, which corresponds to the general formula
\be
T_{\mu\nu}(x)= \frac 2 {\sqrt{g}} \frac {\delta S}{\delta g^{\mu\nu}}\label{Tmunudef}
\ee
and reads
\be
\widehat T_{\mu\nu}= \frac i4 \overline {\psi} \gamma_\mu
{\stackrel{\leftrightarrow}{\nabla_\nu}}\psi+(\mu\leftrightarrow\nu)- g_{\mu\nu}\frac i2  \overline \psi \gamma^\lambda {\stackrel{\leftrightarrow}{\nabla_\lambda}}\psi=T_{\mu\nu} - g_{\mu\nu } T_\lambda{}^\lambda\label{emtmodified}
\ee
This ambiguity in the definition of the em tensor gives rise to an ambiguity in the definition of the trace anomaly.  Such an uncertainty is in fact resolved by the definition  \eqref{Duff}: thanks to the latter, the second term of $\widehat T_{\mu\nu}$ drops out.
This will be verified later on. For the time being we proceed with \eqref{emt}, which is simpler.

The anomaly we are after can only appear at one-loop in the trace of the em tensor. Therefore we have to compute the correlators that contain one insertion of the latter plus insertions of the vector currents. With reference to the definition \eqref{Duff}, at the lowest order we have two possibilities:  the correlator $ \eta^{\mu\nu}\langle 0|{\cal T}T_{\mu\nu}(x) j_{\lambda}(y) j_{\rho}(z)|0\rangle$ and the correlator $\langle 0|{\cal T}T_{\mu}^\mu(x) j_{\lambda}(y) j_{\rho}(z)|0\rangle$. The first means that we regularize and compute the correlator $\langle 0|{\cal T}T_{\mu\nu}(x) j_{\lambda}(y) j_{\rho}(z)|0\rangle$ and afterwards we contract the two indices $\mu$ and $\nu$. The second means that we consider the correlator of $T_\mu^\mu$ inserted at $x$ and two currents at $y$ and $z$, regularize and compute it. Generally speaking, the two procedures leads to different results. Notice that
$T_\mu^\mu(x)$ can vanish on shell in conformal invariant theories, but in general does not vanish off-shell. {An important remark is that} the trace $T_\mu^\mu(x)$ is an irreducible component of $T_{\mu\nu}(x)$. So,  the two above-mentioned amplitudes, when regulated, are generally different. This situation is to be contrasted with the case of the three-point current amplitude $\langle 0|{\cal T}j_{\mu}(x) j_{\lambda}(y) j_{\rho}(z)|0\rangle$, whose divergence (see Appendix) is the same as  $\langle 0|{\cal T}\partial^\mu j_{\mu}(x) j_{\lambda}(y) j_{\rho}(z)|0\rangle$. In this case there is no regularization ambiguity.

Hereafter we shall use the definition \eqref{Duff} for the trace anomaly in the sense just explained. So to the lowest order\footnote{One should consider also the lower order amplitude $\langle t \,j\rangle$, but it is easy to show that
\be 
\langle 0|{\cal T} T_\mu^\mu (x) j_\lambda(y) |0\rangle = \eta^{\mu\nu} \langle 0|{\cal T}  T_{\mu\nu} (x) j_\lambda(y) |0\rangle=0\0
\ee} in the perturbative approach the trace anomaly will be given by:
\be
\langle\!\langle T_\mu^\mu(x)\rangle\!\rangle = \eta^{\mu\nu} \langle 0|{\cal T}T_{\mu\nu}(x) j_{\lambda}(y) j_{\rho}(z)|0\rangle- \langle 0|{\cal T}T_{\mu}^\mu(x) j_{\lambda}(y) j_{\rho}(z)|0\rangle\label{lowestordertraceanomaly}
\ee

In the sequel we compute these two amplitudes by means of Feynman diagrams. The Feynman rule for the vector-fermion-fermion vertex $V_{vff}$ is $\gamma_\mu$ and the graviton-fermion-fermion vertex $V_{hff}$ is
\be
-\frac i{8} \left[(p-p')_\mu \gamma_\nu + (p-p')_\nu \gamma_\mu\right] \label{2f1g}
\ee
There are also other vertices, but they are not relevant to the present calculation. The fermion propagator is $\frac i {\slashed {p}}$. In order to regularize the Feynman integrals (and only to that purpose) we will add, to the ordinary 3+1, $\delta$ additional dimensions. The conventions for the gamma matrix traces (in Minkowski background) are
\be
\tr (\gamma_\mu\gamma_\nu)= 2^{2+\frac {\delta}2} \eta_{\mu\nu}, \quad\quad
\tr(\gamma_\mu\gamma_\nu\gamma_\lambda\gamma_\rho\gamma_5) =-i\, 2^{2+\frac {\delta}2} \epsilon_{\mu\nu\lambda\rho}\label{gammamatrixtr}
\ee

\subsection{The first correlator}

We start by evaluating the second term in the RHS of \eqref{lowestordertraceanomaly}, i.e. the amplitude schematically denoted $\langle t\, j\, j\rangle$, where $t$ represents the trace of the em tensor,
\be
&&\widetilde F^{(tjj)}_{\lambda\rho}(k_1,k_2) = \frac 12
 \int
\frac{d^4p }{(2\pi)^{4}}\mathrm{tr}\left\{\frac 1{\slashed{p}}     
\gamma_\lambda\frac 1{\slashed{p}
-\slashed{k}_1}\gamma_{\rho}
\frac 1{\slashed{p}-\slashed{q}}(2\slashed{p}-\slashed{q}) \right\}\0\\
& &\stackrel {reg}{=}\frac 12 \int \frac{d^4p d^\delta\ell}{(2\pi)^{4+\delta}}\mathrm{tr}\Big\{\frac{\slashed{p}+\slashed{\ell}}{{p}^2-{\ell}^2}     
\gamma_\lambda\frac{\slashed{p}-\slashed{k}_1+\slashed{\ell}}{(p-k_1)^2 -\ell^2} \gamma_{\rho}+ \gamma_\lambda\frac{\slashed{p}-\slashed{k}_1+\slashed{\ell}}{(p-k_1)^2 -\ell^2} \gamma_{\rho}\frac{\slashed{p}-\slashed{q}+\slashed{\ell}}{(p-q)^2-\ell^2} \Big\}\label{triangletjj}
\ee
In the second line we have regularized the integral by introducing an additional momentum $\ell_\mu$, with $\mu= 4,\ldots,\delta-1$. Then we have rewritten $2\slashed{p}-\slashed{q}$ as $ \slashed{p} +\slashed{\ell} +\slashed{p}-\slashed{q}+\slashed{\ell}$ and simplified. 

Adding the cross term $k_1\leftrightarrow k_2, \lambda\leftrightarrow \rho$  we get
\be
&&\widetilde F^{(tjj)}_{\lambda\rho}(k_1,k_2) +\widetilde F^{(tjj)}_{\rho\lambda}(k_2,k_1) \0\\
&&= 4\times 2^{\frac \delta 2 } \int \frac{d^4p d^\delta\ell}{(2\pi)^{4+\delta}}\left\{ \frac{\ell^2 \eta_{\lambda\rho} +p_\lambda (p-k_1)_\rho -\eta_{\lambda\rho}\, p\!\cdot\! (p-k_1) +(p-k_1)_\lambda p_\rho} {(p^2 -\ell^2) ((p-k_1)^2 -\ell^2)}\right.\0\\
&&\quad\quad\quad\quad\quad\quad\quad\quad\quad+\left. \frac{\ell^2 \eta_{\lambda\rho} +p_\lambda (p-k_2)_\rho -\eta_{\lambda\rho}\, p\!\cdot\! (p-k_2) +(p-k_2)_\lambda p_\rho} {(p^2 -\ell^2) ((p-k_2)^2 -\ell^2)}\right\}
\label{trianglettjj1}
\ee
Let us deal first with the first line of the last integral. Introducing a Feynman parameter $x$ and  shifting $p\to p+k_1$  and Wick rotating ($p^0\to ip^0$) one gets\footnote{In order to avoid clogging our formulas with the Euclidean label such as $p^E,...$, we keep the same symbols for the Wick-rotated momenta $p,k_1,k_2$ and $q$, with the only change represented by the momentum square acquiring the opposite sign.}
\be
 &=& 4 i \times 2^{\frac \delta 2 }
 \int \frac{d^4p }{(2\pi)^{4}}\int_0^1 dx \int \frac{d^4p d^\delta\ell}{(2\pi)^{4+\delta}}\frac {\left(\ell^2+ \frac {p^2}2\right)
 \eta_{\lambda\rho}+x(x-1) \left(2 k_{1\lambda} k_{1\rho} + \eta_{\lambda\rho} k_1^2\right)}
{\left(p^2 +\ell^2 +x(1-x)k_1^2\right)^2}\0\\
&=&\frac i{6\pi^2} \left( k_{1\lambda} k_{1\rho}  + \eta_{\lambda\rho} k_1^2\right) \left( \frac 2{\delta} +\gamma -\frac 53 +\log \frac {k_1^2}{2\pi}\right)\0
\ee
Adding also the second line we get
\be
\widetilde F^{(tjj)}_{\lambda\rho}(k_1,k_2)+\widetilde F^{(tjj)}_{\rho\lambda}(k_2,k_1)&=& \frac i{12\pi^2} \left( k_{1\lambda} k_{1\rho}  + \eta_{\lambda\rho} k_1^2\right) \left( \frac 2{\delta} +\gamma -\frac 53 +\log \frac {k_1^2}{2\pi}\right)\0\\
&&+ \frac i{12\pi^2} \left( k_{2\lambda} k_{2\rho}  + \eta_{\lambda\rho} k_2^2\right) \left( \frac 2{\delta} +\gamma -\frac 53 +\log \frac {k_2^2}{2\pi}\right)\label{triangletVV1}
\ee
This contains a non-local part. It is a (so-called) semi-local term which is necessary in order to  satisfy the conformal Ward identity.

\subsection{The conformal Ward identity} 
\label{ss:confWI}

To find the WI's for our case we have to start from the effective action that include both the vector current and the  energy-momentum tensor, which is born out of the action \eqref{actionS},
\be
W[h,V]&=&W[0]  + \sum_{n,r=1} ^\infty \frac { i^{n+r-1}}{{2^{n}}n!r!} \int
\prod_{i=1}^n
dx_i\sqrt{g(x_i)} h^{\mu_i\nu_i}(x_i) \prod_{l=1}^r d y_l\sqrt{g(y_l)}  e_{a_l}^{\lambda_l}(y_l) V_{\lambda_l}(y_l)\0  \\
&&\quad\quad\quad\cdot \langle 0| {\cal T}\widehat T_{\mu_1\nu_1}(x_1)\ldots \widehat T_{\mu_n\nu_n}(x_n)j^{a_1}(y_1) \ldots j^{a_r}(y_r)|0\rangle\label{WhV}
\ee
The introduction of the vierbein is dictated by the coupling between the current and the gauge potential in the presence of a nontrivial background metric: $V_\lambda(x) \bar \psi(x) e_a^\lambda(x) \gamma^a \psi(x)$. Notice that in this definition we have used $\widehat T_{\mu\nu}$; this is because, in the subsequent manipulations, we will refer to \eqref{emtmodified}.
A remark is in order about the coefficients of this expansion. They must be consistent with the definition of graviton emission vertices. Remember that $\frac {\delta S}{\delta g{\mu\nu}}\vert_{g=\eta}= \frac 12 \widehat T_{\mu\nu}$. This explain the factor $2^n$ in the denominator of \eqref{WhV}.

Differentiating with respect to $h_{\mu\nu}(x)$  and setting $h=0$ we obtain $\frac 12\langle\! \langle \widehat T_{\mu\nu}(x)\rangle\!\rangle$. Therefore
\be
\left. 2\frac {\delta W}{\delta h_{\mu\nu}(x)}\right\vert_{h=0}&=&i \int  d^4y_2\, \eta_{\mu\nu}\, V_{\lambda_1}(x)V_{\lambda_2}(y_2)\delta^{\lambda_1}_{a_1} \delta^{\lambda_2}_{a_2}
\langle 0|{\cal T} j^{a_1}(x) j^{a_2}(y_2)|0\rangle\0\\
&&-\frac i2\int  d^4y_2 V_{\lambda_1}(x)V_{\lambda_2}(y_2)\left(\delta^{\lambda_1}_\mu \eta_{a_1\nu}+ \delta^{\lambda_1}_\nu \eta_{a_1\mu}\right)\langle 0|{\cal T} j^{a_1}(x) j^{a_2}(y_2)|0\rangle\0\\
&&  -\frac 12\int d^4y_1 d^4y_2 V^{\lambda_1}(y_1)V^{\lambda_2}(y_2) \langle 0|{\cal T}\widehat T_{\mu\nu}(x)j_{\lambda_1}(y_1) j_{\lambda_2}(y_2)|0\rangle\label{WVh1}
\ee
Next differentiating with respect to $V^\lambda (y)$ and $V^{\rho}(z)$ and setting $V_\lambda=0$ we get
\be
\left. \frac {\delta^2\langle\! \langle \widehat T_{\mu\nu}(x)\rangle\!\rangle}{\delta V^\lambda(y) \delta V^\rho(z)}\right\vert_{V=0}&=& i\eta_{\mu\nu} \bigl(\delta(x-y)  \langle 0|{\cal T} j_\lambda(x) j_\rho(z) |0\rangle+\delta(x-z)  \langle 0|{\cal T} j_\rho(x) j_\lambda(y) |0\rangle\bigr)\0\\
&& -\frac i2\, \eta_{\mu\lambda} \,\delta(x-y) \langle 0|{\cal T} j_\nu(x) j_\rho(z) |0\rangle-\frac i2 \,\eta_{\mu\rho}\,\delta(x-z) \langle 0|{\cal T} j_\nu(x) j_\lambda(y) |0\rangle\0\\
&&-\frac i2\, \eta_{\nu\lambda}\, \delta(x-y) \langle 0|{\cal T} j_\mu(x) j_\rho(z) |0\rangle-\frac i2\, \eta_{\nu\rho}\,\delta(x-z) \langle 0|{\cal T} j_\mu(x) j_\lambda(y) |0\rangle\0\\
&& - \langle 0| {\cal T}\widehat T_{\mu\nu}(x) j_\lambda(y) j_\rho(z) |0\rangle\label{tjjlevel2}
\ee
Now we saturate this relation with $\eta^{\mu\nu}$ and notice that
$\widehat T_\mu{}^\mu=-3 T_\mu{}^\mu$. It is easy to see that the numerical factor 3 factors out  and can be dropped from the equation obtained by equating the RHS of \eqref{tjjlevel2} to zero. Thus we obtain 
\be
 \langle 0| {\cal T} T_{\mu}^\mu (x) j_\lambda(y) j_\rho(z) |0\rangle+  i \Big(\delta(x-y) \langle 0|{\cal T} j_\lambda(x) j_\rho(z) |0\rangle+\delta(x-z) \langle 0|{\cal T} j_\lambda(y) j_\rho(x) |0\rangle\Big)=0\label{WItjj}
\ee
This is the appropriate conformal WI  for a fermionic system coupled to an external gauge potential $V_\mu$, at level 2, in absence of anomalies. Fourier transforming and Wick rotating it (under a Wick rotation the two-point function changes sign, the three-point function gets multiplied by $-i$) we get
\be
-i   \left( \widetilde F_{\lambda\rho}(k_1,k_2) +\widetilde F_{\rho\lambda}(k_2,k_1)+  \langle 0|{\cal T}\tilde j_\lambda (k_1) \tilde j_\rho (-k_1)|0\rangle + \langle 0|{\cal T}\tilde j_\lambda (k_2) \tilde j_\rho (-k_2)|0\rangle \right)=0\label{WItjjFT}
\ee
multiplied by $\delta(q-k_1-k_2)$. This is the WI if no anomaly is present.

\subsection{The 2-point current correlator}

In order to verify \eqref{WItjj} (or \eqref{WItjjFT}) we need the two-point current correlators.
The 2-current correlator is given by a bubble diagram with a fermion propagating in the internal lines
and two gluons, one ingoing and the other outgoing, with the same momentum
\be
\langle 0|{\cal T}\tilde j_\lambda (k) \tilde j_\rho (-k)|0\rangle&=&  \int
\frac{d^4p }{(2\pi)^{4}}\mathrm{tr}\left\{\frac 1{\slashed{p}}     
\gamma_\lambda \frac 1{\slashed{p}
-\slashed{k}}\gamma_{\rho} \right\}=\int
\frac{d^4pd^\delta \ell }{(2\pi)^{4+\delta}}\mathrm{tr}\left\{\frac 1{\slashed{p}+\slashed{\ell}}     
\gamma_\lambda \frac 1{\slashed{p}
-\slashed{k}+\slashed{\ell}}\gamma_{\rho} \right\}\label{2ptjj}
\ee
This is equal to the first term in the third line of \eqref{triangletjj}, apart from the coefficient $\frac 12$. Therefore 
\be
\langle 0|{\cal T}\tilde j_\lambda (k) \tilde j_\rho (-k)|0\rangle=-\frac i{12\pi^2} \left( k_{\lambda} k_{\rho}  + \eta_{\lambda\rho} k^2\right) \left( \frac 2{\delta} +\gamma -\frac 53 +\log \frac {k^2}{2\pi}\right)\label{2ptjj1}
\ee
in Euclidean background.
From this one can see that the WI \eqref{WItjjFT} is satisfied.

\subsection{The second correlator}

As we have seen, replacing \eqref{triangletVV1} and \eqref{2ptjj1} in \eqref{WItjjFT} we see that the  WI is satisfied. One could conclude therefore that in this case there is no anomaly. However \eqref{triangletVV1} corresponds to $\langle\!\langle g^{\mu\nu} T_{\mu\nu}(x) \rangle\!\rangle$, and, comparing it with $g^{\mu\nu}\langle\!\langle T_{\mu\nu}(x) \rangle\!\rangle $, whose relevant amplitude  is
\be
&&\widetilde{\widetilde F}^{(tjj)}_{\lambda\rho}(k_1,k_2) = \frac 12
 \int
\frac{d^4p d^\delta\ell}{(2\pi)^{4+\delta}}\mathrm{tr}\left\{\frac 1{\slashed{p}+\slashed{\ell}}     
\gamma_\lambda\frac 1{\slashed{p}
-\slashed{k}_1+\slashed{\ell}}\gamma_{\rho}
\frac 1{\slashed{p}-\slashed{q}+\slashed{\ell}}(2\slashed{p}-\slashed{q}) \right\},\label{doubleFtjj}
\ee
one finds a difference
\be
\Delta{\widetilde F}^{(tjj)}_{\lambda\rho}(k_1,k_2)& =& \frac 12
 \int
\frac{d^4pd^\delta \ell }{(2\pi)^{4+\delta}}\mathrm{tr}\left\{\frac 1{\slashed{p}+\slashed{\ell}}     
\gamma_\lambda\frac 1{\slashed{p}
-\slashed{k}_1+\slashed{\ell}}\gamma_{\rho}
\frac 1{\slashed{p}-\slashed{q}+\slashed{\ell}}2\slashed{\ell} \right\},\label{DeltaFtjj}
\ee
which is a local term.  Adding the cross term, after a Wick rotation,  one finds
\be
\Delta{\widetilde F}^{(tjj)}_{\lambda\rho}(k_1,k_2)+
\Delta{\widetilde F}^{(tjj)}_{\rho\lambda }(k_2,k_1)=
-\frac i{6\pi^2}\bigl[ \eta_{\lambda\rho}k_1\!\cdot\!k_2+k_{2\lambda}k_{1\rho} \bigr]\label{DeltaFtjjtotal}
\ee
Beware: here the metric is Euclidean!

The RHS is a local term which violates the conformal WI and corresponds to an anomaly.
The definition of the effective energy-momentum tensor is
\be
T_{\mu\nu} = \frac 2{\sqrt{g}} \frac {\delta W}{\delta g^{\mu\nu}},\label{defem}
\ee
The Weyl variation of $g^{\mu\nu}$ is $\delta g^{\mu\nu} = - 2 \omega g^{\mu\nu}$. Therefore
\be
\delta_\omega W = \int d^4x \, \frac {\delta W}{\delta g^{\mu\nu}} = - \int d^4x\, \sqrt{g}\, \omega \, \langle\!\langle T_\mu^\mu\rangle \!\rangle\label{deltaomegaW}
\ee
which classically vanishes. We define the integrated anomaly as follows
\be
{\cal A}_\omega =\int d^4x \sqrt{g}\, \omega \, \left(g^{\mu\nu} \langle\!\langle T_{\mu\nu} \rangle\!\rangle-  \langle\!\langle T_\mu^\mu\rangle \!\rangle\right)\label{trueAomega}
\ee
We have already remarked that if we use the definition  \eqref{emtmodified} instead of \eqref{emt}, the second piece of the former drops out in eq.\eqref{trueAomega}.

In order to obtain the form of the anomaly in coordinate representation we proceed as follows. We have in general
\be
&&-\frac 12 \int d^4y d^4z \, V^\lambda (y) V^\rho(z) \langle 0| T_\mu^\mu (x) j_\lambda(y) j_\rho(z) |0\rangle
= -\frac 12 \int d^4y d^4z \, V^\lambda (y) V^\rho(z)\0\\
&&\quad\quad \cdot \int \frac{ d^4q}{(2\pi)^4}  \frac{ d^4k_1}{(2\pi)^4}  \frac{ d^4k_2}{(2\pi)^4}e^{-iq\cdot  x +i k_2\cdot y+ik_2\cdot z} \delta^{(4)}(q-k_1-k_2)  
\langle 0|\tilde T_\mu^\mu (q) \tilde j_\lambda(-k_1) \tilde j_\rho(k_2) |0\rangle\0
\ee 
Inserting \eqref{DeltaFtjjtotal} in the RHS gives the contribution corresponding to the anomaly in coordinates:
\be
&&-\frac 12 \int d^4y d^4z \, V^\lambda (y) V^\rho(z)
 \int \frac{ d^4q}{(2\pi)^4}  \frac{ d^4k_1}{(2\pi)^4}  \frac{ d^4k_2}{(2\pi)^4}e^{-iq\cdot  x +i k_1\cdot y+ik_2\cdot z} \delta^{(4)}(q-k_1-k_2)  \0\\
&&\cdot \left(-\frac i{6\pi^2}\big[ \eta_{\lambda\rho}k_1\!\cdot\!k_2+ k_{2\lambda}k_{1\rho} \big]\right)\label{Aomega0}\\
&=&
 \frac {i}{12\pi^2}  \int d^4y d^4z \, V^\lambda (y) V^\rho(z)
 \big(-\eta_{\lambda\rho}\partial_x\!\cdot \! \partial_z- \partial_\rho^y \partial_\lambda^z\big) \delta^{(4)}(x-y) \delta^{(4)}(x-z)\0\\
&=& -\frac i{12\pi^2} \big(\partial_\lambda V_\rho \partial^\lambda V^\rho + \partial_\lambda V_\rho \partial^\rho V^\lambda\big)\0
\ee
Therefore, after migrating back to Minkowski,  the trace anomaly at the (non-trivial) lowest level of approximation is
\be
{\cal A}_\omega =-\frac 1{12\pi^2}  \int d^4x\, \omega \, \left(-\partial_\nu V_\lambda \partial^\nu V^\lambda + \, \partial_\nu V_\lambda \partial^\lambda V^\nu\right)=\frac 1{24\pi^2}  \int d^4x\, \omega \, F_{\nu\lambda }F^{\nu\lambda}\label{trueAomega1}
\ee
This is a (trivially) consistent conformal anomaly because in perturbative cohomology (see Appendix B) at the lowest order, we have
\be
\delta_\omega^{(0)}\omega=0, \quad\quad \delta_\omega^{(0)} V_\mu =0 \label{delta0V}
\ee
so
\be
\delta^{(0)}_\omega {\cal A}_\omega=0\label{deltaomegaA}
\ee
It is clear that the all-order expression of this anomaly is
\be
{\cal A}_\omega =\frac 1{24\pi^2}  \int d^4x\,\sqrt{g}\, \omega \, F_{\nu\lambda }F^{\nu\lambda},\label{trueAomega2}
\ee
which is invariant both under diffeomorphisms and under gauge transformations. When diffeomorphisms are involved, however, one must pay double attention. It is usually assumed that diffeomorphisms are conserved in 4d. This  is due to the fact that consistent chiral diffeomorphism anomalies are uniquely linked to the existence of the third order symmetric adjoint invariant tensor of the relevant Lie algebra, which,  in 4d, is the Lie algebra of $GL(3,1)$ and, thus, the relevant tensor vanishes. But here we are considering the possibility that diffeomorphisms are violated by the coupling of a Dirac fermion to a vector potential, and an anomaly proportional to $\int d^4x\, \partial \!\cdot\! \xi\, F_{\mu\nu}F^{\mu\nu}$ (where $\xi^\mu$ is the general coordinate transformation parameter) is consistent and algebraically not excluded. To cancel it a counterterm $\int d^4x \, h\, F_{\mu\nu}F^{\mu\nu}$ would be necessary. Since $h=h^\mu_\mu$ transforms as $\delta_\omega h= 8\omega$ under Weyl transformations, such a counterterm  would generate an anomaly of the same kind as \eqref{trueAomega1} and, thus, modify its coefficient. It is therefore important to verify that such a diffeomorphism anomaly generated by the coupling to a gauge field $V_\mu$ is absent in our theory.
 This is what we intend to show next.

\subsection{Diffeomorphisms are conserved}

We have to show that the WI for diffeomorphisms is respected when coupling our Dirac fermion to $V_{\mu}$. In order to derive the relevant WI we return to subsection \ref{ss:confWI} and, precisely, to eq.\eqref{tjjlevel2}. We differentiate the RHS of the latter with respect to $x^\mu$, and equate it to zero. The WI we obtain is formulated in terms of $\widehat T_{\mu\nu}$. The identity simplifies considerably if we express it in terms of $T_{\mu\nu}$, for, using \eqref{WItjj}, the first line in the RHS of \eqref{tjjlevel2} drops out. Therefore we are left with
\be
0&=& -\frac i2\,\Big( \partial_\lambda ^x\left(\,\delta(x-y) \langle 0|{\cal T} j_\nu(x) j_\rho(z) |0\rangle\right)+\partial_\rho^x\left(\delta(x-z) \langle 0|{\cal T} j_\nu(x) j_\lambda(y) |0\rangle\right)\Big)\0\\
&&-\frac i2\,\Big( \eta_{\nu\lambda}\,\partial^\mu_x\left( \delta(x-y) \langle 0|{\cal T} j_\mu(x) j_\rho(z) |0\rangle\right)+ \, \eta_{\nu\rho}\,\partial^\mu_x\left(\delta(x-z) \langle 0|{\cal T} j_\mu(x) j_\lambda(y) |0\rangle\right)\Big)\0\\
&& - \partial^\mu_x \langle 0| {\cal T} T_{\mu\nu}(x) j_\lambda(y) j_\rho(z) |0\rangle\label{diffWI}
\ee
If we denote by $\widetilde T_{\mu\nu\lambda\rho} (k_1,k_2)$  the Fourier transform of 
$\langle 0| {\cal T}T_{\mu\nu}(x) j_\lambda(y) j_\rho(z) |0\rangle$, the Fourier transform of eq.\eqref{diffWI} is
\be
-i q^\mu \left( \widetilde T_{\mu\nu\lambda\rho} (k_1,k_2)+  \widetilde T_{\mu\nu\rho\lambda} (k_2,k_1)\right)\!\!\!
& =&\!\!\! \frac 12 \Big(q_\rho \langle 0|{\cal T}\tilde j_\nu (k_1) \tilde j_\lambda (-k_1)|0\rangle +q_\lambda \langle 0|{\cal T}\tilde j_\nu (k_2) \tilde j_\rho (-k_2)|0\rangle \0\\
+\eta_{\nu\rho} q^\mu\!\!\!\!\!\!\!\!
&&\!\!\!\!\!\!\!\! \langle 0|{\cal T}\tilde j_\mu (k_1) \tilde j_\lambda (-k_1)|0\rangle +\eta_{\nu\lambda}q^\mu \langle 0|{\cal T}\tilde j_\mu (k_2) \tilde j_\rho (-k_2)|0\rangle\Big)\label{FdiffWI}
\ee
where $q=k_1+k_2$. Thus we have to compute $q^\mu \widetilde T_{\mu\nu\lambda\rho}(k_1,k_2)$:
\be
&&q^\mu \widetilde T_{\mu\nu\lambda\rho}(k_1,k_2)=\frac 12  \int
\frac{d^4p}{(2\pi)^4}\mathrm{tr}\left[\frac 1{\slashed{p}} \gamma_\lambda \frac 1 {\slashed{p}-\slashed{k}_1}\gamma_\rho\frac 1{\slashed{p}-\slashed{q}} \left(q\!\cdot\!(2p-q)\gamma_\nu +(2p-q)_\nu \slashed{q}\right) \right]\0\\
&\stackrel{reg}{=}& \frac 12  \int
\frac{d^4pd^\delta\ell}{(2\pi)^{4+\delta}}\mathrm{tr}\left[\frac 1{\slashed{p}+\slashed{\ell}} \gamma_\lambda \frac 1 {\slashed{p}-\slashed{k}_1+\slashed{\ell}}\gamma_\rho\frac 1{\slashed{p}-\slashed{q}+\slashed{\ell}} \left(q\!\cdot\!(2p-q) \gamma_\nu+(2p-q)_\nu \slashed{q}\right) \right]\label{gauge&diff0}
\ee

Using now $\slashed{q} = \slashed {p} +\slashed{\ell} -(\slashed{p} -\slashed{q} +\slashed {\ell})$, and $q\!\cdot\!(2p-q)= p^2 +\ell^2 -((p-q)^2+\ell^2)$ we obtain an easier-to-deal-with expression of the integrand
\be
q^\mu \widetilde T_{\mu\nu\lambda\rho}(k_1,k_2)&=&\frac 12  \int
\frac{d^4pd^\delta\ell}{(2\pi)^{4+\delta}}\Bigg[\mathrm{tr}\Big( (\slashed {p}+\slashed{\ell}) \gamma_\lambda (\slashed {p}-\slashed {k}_1+\slashed{\ell})\gamma_\rho (\slashed {p}-\slashed {q}+\slashed{\ell})\gamma_\nu\Big)\label{qmuTmunulr}\\
&&\quad\quad\quad\cdot\left( 
\frac 1{((p-k_1)^2-\ell^2)((p-q)^2-\ell^2)}-\frac 1{(p^2-\ell^2)((p-k_1)^2-\ell^2)} \right)\0\\
&&+ \mathrm{tr}\left( \gamma_\lambda   \frac 1 {\slashed{p}-\slashed{k}_1+\slashed{\ell}}\gamma_\rho\frac 1{\slashed{p}-\slashed{q}+\slashed{\ell}}- \frac 1{\slashed{p}+\slashed{\ell}} \gamma_\lambda \frac 1 {\slashed{p}-\slashed{k}_1+\slashed{\ell}}\gamma_\rho\right)(2p_\nu -q_\nu)\Bigg]\0
\ee
This expression contains four integrations, but two of them are copies of the other two. To see it one can proceed by changing variables as follows: $p\to p+q$, followed by $p\to -p$, then $k_1 \leftrightarrow k_2, \lambda\leftrightarrow \rho$; finally using the invariance of the trace under transposition.

Next, selecting two independent terms, one works out the gamma matrix algebra, introduces a Feynman parameter and performs a Wick rotation, after which the integrals can be easily calculated. After some algebra the final result is
\be
q^\mu \widetilde T_{\mu\nu\lambda\rho}(k_1,k_2)\!\!\!&=&\!\!\!\frac i{24\pi^2} \left( k_{1\lambda} k_{1\nu} q_\rho-\eta_{\lambda\rho} k_1^2 q_\rho +\eta_{\nu\rho} ( k_1^2 k_{2\lambda} -k_1\!\cdot\! k_2\, k_{1\lambda})\right)\left( \frac 2{\delta} +\gamma-\frac 53 +\log \frac {k_1^2}{2\pi}\right)\0\\
&&\label{divergenceTmunu}
\ee 
to which we have to add the cross term with $k_1\leftrightarrow k_2, \lambda\leftrightarrow \rho$. As a consequence we are left with two semi-local terms, which are exactly what is needed in order to satisfy the WI \eqref{FdiffWI}. 

In conclusion the WI for diffeomorphims is satisfied and there is no anomaly\footnote{On the contrary, if we replace Dirac fermions with Weyl fermions this result is not guaranteed.}. The trace anomaly \eqref{trueAomega2} is therefore confirmed.

\section{The Seeley-DeWitt approach}

In this section we apply the Seeley-DeWitt approach (see \cite{DeWitt}-\cite{DeWittglobal}) to the same theory of Dirac fermions coupled to a vector potential $V_\mu$, \eqref{actionS}, where the covariant operator
\be
\nabla_\mu = D_\mu +\frac 12   \omega_\mu + V_\mu \label{nablamu}
\ee
features.

To apply the SDW method we need the square of the Dirac operator, which is not selfadjoint
\be
\left({\slashed{\nabla}}^2\right)^\dagger= \gamma_0 {\slashed{\nabla}}^2\gamma_0\label{D2dagger}
\ee
To get a self-adjoint operator we apply a Wick rotation, which means: $x^0\to \tilde x^0= -i x^0, k^0\to \tilde k^0=i k^0$ and $\gamma^0 \to \tilde \gamma^0=- i\gamma^0$, while $x^i, k_i, \gamma^i$  remain unchanged. From now on a tilde represents a Wick rotated object. In particular we have
\be
\left(\widetilde{\slashed{\nabla}}^2\right)^\dagger= \widetilde{\slashed{\nabla}}^2\label{widetildeDsquare}
\ee
Therefore we can use $\widetilde{\slashed{\nabla}}^2$ for the SDW approach and return to the Minkowski metric with a reverse Wick rotation after applying this method. However, as before, we will avoid clogging the formulas with Euclidean symbols. We will use all the time Minkowski symbols, but we will leave the gamma matrix products indicated, without using the gamma matrix algebra to simplify them, before returning to the Lorentz metric with an inverse Wick rotation. This amounts, in practice, to using all the time non-Wick rotated quantities, replacing them in the relations appropriate to their Euclidean counterparts. This is what we will do in the sequel.

With this attitude in mind we define  the amplitude
\be
\langle   x, s|  x',0\rangle = \langle  x| e^{i 
\EF
 s}| x'\rangle\label{hatampA}
\ee
which satisfies the (heat kernel) differential equation
\be
i \frac {\partial }{\partial  s} \langle  x, s| x',0\rangle= - \EF_{ x} \langle  x, s|  x',0\rangle
\equiv K( x,  x', s) \label{hatdiffeqforA}
\ee
where $\EF_{ x}$ is the above-mentioned quadratic differential operator
\be
\EF_{ x}=\nabla_\mu { g}^{\mu\nu}
\nabla_\nu- \frac
14  R+{\cal V}\label{hatEHx}
\ee
where $ {\cal V}= \Sigma^{ab} \EV_{ab}= \Sigma^{ab}\, e_a^\mu  e_b^\nu\,  V_{\mu\nu} $, and $F_{\mu\nu} $ is the curvature of
$V_\mu$.
Then we make the ansatz
\be
\langle  x, s|  x',0\rangle = -\lim_{m\to 0}\frac i{16\pi^2}\frac {\sqrt{D( x, x')}} { s^2}
e^{i\left(\frac {\sigma( x, x')} {2s}-m^2 s\right)}\Phi( x, x', s)\label{ansatz3}
\ee
where $ D( x, x')$ is the VVM determinant and
$\sigma(x,x')$ is the world function. $\Phi( x, x', s)$ is a
function to be determined. It is useful to introduce also the mass parameter
$m$, which
we will eventually set to zero.
In the limit $  s\to 0$ the RHS of \eqref{ansatz3} becomes the
definition
of a delta function multiplied by $\Phi$. More precisely,  since it must
be $\langle x,0|  x',0\rangle=\delta( x, x')$,
and
\be
\lim_{ s\to 0} \frac i{16\pi^2} \frac {\sqrt{ D(
x, x')}} { s^2}
\, e^{i\left(\frac {\sigma( x, x')} {2
s}-m^2 s\right)}
= \sqrt{| g( x)|}\,\,
\delta( x, x'),\label{hatlimdelta}
\ee
we must have
\be
\lim_{ s\to 0} \Phi( x, x', s)={\bf
1}\label{hatlimitPhi}
\ee
Eq.\eqref{hatdiffeqforA} becomes an equation for $ \Phi(
x, x', s)$. After
some algebra one gets
\be
i\frac {\partial \Phi}{\partial  s} +\frac i{ s}
 \nabla^\mu
 \Phi  \nabla_\mu \sigma
+\frac 1{\sqrt{ D}}  \nabla^\mu  \nabla_\mu
\left(\sqrt{ D}  \Phi\right)-\left(\frac 14 
R- {\cal V}-m^2\right)\Phi=0\label{hateqforPhi}
\ee
Now we expand
\be
\Phi( x, x', s)
= \sum_{n=0}^\infty  a_n( x, x') (i s)^n
\label{hatPhiexp}
\ee
with the boundary condition $[ a_0]=1$. The $ a_n$ must satisfy
the recursive relations:
\be
(n+1) a_{n+1} +  \nabla^\mu  a_{n+1}  \nabla_\mu
 \sigma - \frac 1{\sqrt{ D}}  \nabla^\mu 
\nabla_\mu
\left(\sqrt{ D}   a_n\right) +\left(\frac 14  R- {\cal V}
-m^2\right)  a_n=0
\label{hatrecursive}
\ee
Using these relations and the coincidence limits it is possible to compute
each coefficient $a_n$ at the coincidence limit. In particular
\be
[ a_2] 
&=& \frac 12 m^4 -\frac 1{12} m^2  R +\frac 1 {288}  R^2 -\frac
1{120}  R_{;\mu}{}^\mu
-\frac 1{180}  R_{\mu\nu} R^{\mu\nu} + \frac 1{180} 
R_{\mu\nu\lambda\rho} R^{\mu\nu\lambda\rho}\label{a2}\\
&&-\frac 12 {\cal V}^2+\frac 16 \nabla^\lambda  \nabla^\lambda\left( {\cal V} \right) +\frac  1{12}
\left(\frac 12 {\cal R}_{\mu\nu}+ V_{\mu\nu}\right)\left(\frac 12 {\cal R}^{\mu\nu}+ V^{\mu\nu}\right)   \0
\ee
where ${\cal R}_{\mu\nu}= { R}_{\mu\nu}{}^{ab}\Sigma_{ab}$.

In this method let us set
\be
 W= -\frac 12 \int_0^\infty \frac {d s}{i s}
e^{i \EF s}+ {\rm const}\equiv
  L+{\rm const}\label{W+const}
\ee
where $ L$ is the relevant  effective action
\be
 L=\int d^d x \,  L( x) \label{Llocal}
\ee
It  can be written  as
\be
 L( x) =  -\frac 12 \tr \int_0^\infty \frac {d
s}{i s}
 K( x, x', s)\label{WK}
\ee
where the kernel $ K$ is defined by
\be
 K( x, x ', s)= e^{i \EF \, s}
\delta( x, x')\label{Kxx'}
\ee
Inserted in $\delta_{\omega} W$, under the symbol $\Tr$, it means
integrating over $x$ after
taking the limit $x'\to x$. So, looking at \eqref{ansatz3}, in dimension $d$,
\be
  K( x, x, s)
=\frac i{(4\pi i  s)^{\frac d2}}\,\sqrt{ g}\, e^{-im^2
s} [
\Phi( x, x, s)]
\label{Kxx}
\ee
From now on a Wick rotation is understood. Continuing analytically in $d$ one finds
\be
 L( x) &=& \frac 1{32\pi^2}\left( \frac 1{d-4} -\frac
34\right)\tr
\left(m^4-2m^2 {[ a_1]}+2 {[ a_2]}\right)\sqrt{
g}\label{Lxd4}\\
&&+  \frac i{64\pi^2}\tr \int_0^{\infty}d s\,  \ln(4\pi i \mu^2 
s)
\sqrt{ g}\frac {\partial^3}{\partial(i s)^3} \left(
e^{-im^2 s} [
\Phi( x, x, s)]\right)\0
\ee
where $\tr$ refers to the gamma matrix trace.
The last line depends explicitly on the parameter $\mu$ and represents a nonlocal
part, which cannot give rise to anomalies. Let us take the variation under a Weyl transformation
\be
\delta_{\omega}\,  g_{\mu\nu}(x) = 2\, \omega(x) \, g_{\mu\nu}(x), \quad\quad
\delta_\omega \sqrt{g} =d\, \omega\, \sqrt{g},\quad\quad \delta_\omega\left( V_{\mu\nu} V^{\mu\nu}\right) =-4\,\omega \,V_{\mu\nu} V^{\mu\nu} \label{ExtWeyl}
\ee
and consider the flat limit $\sqrt{ g} \to 1$ as well as  $m\to 0$. 
Focusing on the $ V_\mu$ dependence in the $d\to 4$ limit we find
\be
\delta_{\omega}\, L( x) =-\frac 1{32\pi^2}\int d^4x \, \frac 43 \omega \,V_{\mu\nu} V^{\mu\nu} \label{deltaomegaL}
\ee
when $ g_{\mu\nu} \to \eta_{\mu\nu}$.
This defines the anomaly
\be
{\cal T}(x)= \frac 1{24\pi^2} V_{\mu\nu} V^{\mu\nu} \label{T(x)}
\ee
which coincides with \eqref{trueAomega1}. { We remark that the - sign and a factor of $\frac 12$ in the coefficient of \eqref{deltaomegaL} comes from the transformation property of $\slashed \nabla$ under a Weyl transformation, which is
\be
\delta_\omega \slashed \nabla=-\frac 12 \{\slashed \nabla, \omega\}\label{Weylnabla}
\ee
This is due to the presence of $\sqrt{g}$ in the action \eqref{actionS} and is the complementary in the action \eqref{actionS} of the transformation property of the `effective' field $\Psi =g^{\frac 14} \psi$, which transforms as
\be
 \delta_{\omega} \Psi=e^{\frac 12 {\omega}} \Psi, \label{PsiWeyl}
\ee
for Weyl transformations. }

We have already pointed out that the result  \eqref{trueAomega1} is the correct one provided the invariance under diffeomorphisms is preserved. In the perturbative case we have verified it by computing the divergence of the em tensor at least to second order in $V_\mu$. In the case of the SDW method this invariance is imbedded in the method itself because the latter is designed to respect the diffeomorphisms.

\section{The case of a right-handed Weyl fermion. Odd parity}

The second example we wish to consider is the case of a Weyl fermion coupled to a vector potential and compute its odd  parity trace anomaly on the basis of the definition \eqref{Duff}. That is we intend now to compute the relation between the (odd parity)  trace $\eta^{\mu\nu}\langle 0|{\cal T}T_{R\mu\nu}(x) j_{R\lambda}(y) j_{R\rho}(z)|0\rangle$ and the (odd parity) correlator $\langle 0|{\cal T}T_{R\mu}^\mu(x) j_{R\lambda}(y) j_{R\rho}(z)|0\rangle$.

Let us start from the following remark. For a right-handed fermion the triangle contribution to the gauge trace anomaly is
\be
\widetilde T^{(R)\mu}_{\mu\lambda\rho}(k_1,k_2) &=&\!\! \frac 12\int
\frac{d^4p}{(2\pi)^4}\mathrm{tr}\left[\frac{1}{\slashed{p}}     
 \gamma_\lambda P_R\frac{1}{\slashed{p}
-\slashed{k}_1} \gamma_{\rho}P_R
\frac{1}{\slashed{p}-\slashed{k}
_1-\slashed{k}_2}(2 \slashed{p}-\slashed{q}) P_R\right]\label{TRjRjR1}\\
&\stackrel{reg}{=}&\!\! \frac 12\int
\frac{d^4pd^\delta\ell}{(2\pi)^{4+\delta}}\mathrm{tr}\left[\frac{1}{\slashed{p}+\slashed{\ell}}     
 \gamma_\lambda P_R \frac{1}{\slashed{p}
-\slashed{k}_1+\slashed{\ell} } \gamma_{\rho}P_R
\frac{1}{\slashed{p}-\slashed{q}+\slashed{\ell}}(2 \slashed{p}+2\slashed{\ell}-\slashed{q})P_R \right]\0\\
&=&\frac 12 \int
\frac{d^4pd^\delta\ell}{(2\pi)^{4+\delta}}\mathrm{tr}\left\{\frac{\slashed{p} }{{p}^2-{\ell}^2}     
 \gamma_\lambda\frac{\slashed{p} 
-\slashed{k}_1}{(p-k_1)^2 -\ell^2} \gamma_{\rho}
\frac{\slashed{p} -\slashed{q}+\slashed{\ell}}{(p-q)^2-\ell^2}(2 \slashed{p}+2\slashed{\ell}-\slashed{q}) P_R\right\}\0
\ee
to which the cross term must be added.

In ref.\cite{BSZ} we have already calculated amplitude for $\langle \partial\!\cdot\! j_R\, j_R\, j_R\rangle$
\be
\widetilde F^{(R,odd)}_{\lambda\rho}(k_1,k_2,\delta)
&=&\left. \int
\frac{d^4pd^\delta\ell}{(2\pi)^{4+\delta}}\mathrm{tr}\left\{\frac{\slashed{p} }{{p}^2-{\ell}^2}     
 \gamma_\lambda\frac{\slashed{p} 
-\slashed{k}_1}{(p-k_1)^2 -\ell^2} \gamma_{\rho}
\frac{\slashed{p} -\slashed{q}}{(p-q)^2-\ell^2}
\slashed{q} P_R\right\}\right\vert_{odd} \0\\
&=& \frac 1{24\pi^2}\epsilon_{\mu\nu\lambda\rho}k_1^\mu k_2^\nu\label{triangledim2'}
\end{eqnarray}
which, multiplied by 2, gives $ \langle 0|{\cal T}\partial^\mu j_{R\mu} (x)  j_{R\lambda}(y) j_{R\rho}(z)|0\rangle$. 

The difference with \eqref{TRjRjR1} apart from the factor $\frac 12$, is the $2 \slashed{p}-\slashed{q}$ factor in the RHS, instead of the $\slashed{q}$ one. On the other hand, let us remark that the odd part of the expression
\be
\Delta\widetilde T^{(R)\mu}_{\mu\lambda\rho}(k_1,k_2) &=&\!\! \frac 12\int
\frac{d^4pd^\delta\ell}{(2\pi)^{4+\delta}}\mathrm{tr}\left[\frac{1}{\slashed{p}+\slashed{\ell}}     
 \gamma_\lambda P_R \frac{1}{\slashed{p}
-\slashed{k}_1+\slashed{\ell} } \gamma_{\rho}P_R
\frac{1}{\slashed{p}-\slashed{q}+\slashed{\ell}}(2 \slashed{p}+2\slashed{\ell}-2\slashed{q})P_R \right]\0\\
&=& \!\! \frac 12 \int
\frac{d^4pd^\delta\ell}{(2\pi)^{4+\delta}}\frac{\mathrm{tr}\left[
\gamma_\lambda {\slashed{p}} 
 \gamma_{\rho}(\slashed{p}-\slashed{k}_1)
P_R\right]}{ (p^2 -\ell^2) ((p-k_1)^2 -\ell^2)}\label{DelatTRJrjR}
\ee
vanishes by symmetry. Now, since $2 \slashed{p}+2\slashed{\ell}-\slashed{q}=2 \slashed{p}+2\slashed{\ell}-2\slashed{q}+ \slashed{q}$, it follows that
\be
\widetilde T^{(R)\mu}_{\mu\lambda\rho}(k_1,k_2)+
\widetilde T^{(R)\mu}_{\mu\rho\lambda}(k_2,k_1)\Big\vert_{odd}=
 \frac 1{24\pi^2}\epsilon_{\mu\nu\lambda\rho}k_1^\mu k_2^\nu,\label{TmumujRjR}
\ee
which is the result for  $\langle 0|{\cal T}T_{R\mu}^\mu(x) j_{R\lambda}(y) j_{R\rho}(z)|0\rangle$.
Now, it is easy to compute 
\be
&&\frac 12\int
\frac{d^4pd^\delta\ell}{(2\pi)^{4+\delta}}\mathrm{tr}\left\{\frac{\slashed{p} }{{p}^2-{\ell}^2}     
 \gamma_\lambda\frac{\slashed{p} 
-\slashed{k}_1}{(p-k_1)^2 -\ell^2} \gamma_{\rho}
\frac{\slashed{p} -\slashed{q}+\slashed{\ell}}{(p-q)^2-\ell^2}
2\slashed{\ell} P_R\right\}\Bigg{\vert}_{odd}\0\\
&=&\frac 1{48\pi^2}\epsilon_{\mu\nu\lambda\rho}k_1^\mu k_2^\nu\label{triangledimtrace1}
\ee
Subtracting \eqref{triangledimtrace1} from \eqref{TmumujRjR} we get
\be
&&\frac 12\int
\frac{d^4pd^\delta\ell}{(2\pi)^{4+\delta}}\mathrm{tr}\left\{\frac{\slashed{p} }{{p}^2-{\ell}^2}     
 \gamma_\lambda\frac{\slashed{p} 
-\slashed{k}_1}{(p-k_1)^2 -\ell^2} \gamma_{\rho}
\frac{\slashed{p} -\slashed{q}}{(p-q)^2-\ell^2}
(2\slashed{p}-\slashed{q} )P_R\right\}\Bigg{\vert}_{odd}+{\it cross}\0\\
&=&  \frac 1{48\pi^2}\epsilon_{\mu\nu\lambda\rho}k_1^\mu k_2^\nu\label{triangledimtrace2}
\ee
which gives  gives the odd parity part of $\eta^{\mu\nu}\langle 0|{\cal T}T_{R\mu\nu}(x) j_{R\lambda}(y) j_{R\rho}(z)|0\rangle$. Therefore, we can say
\be
g^{\mu\nu}\langle\!\langle T_{R\mu\nu}(x) \rangle\!\rangle\Big\vert_{odd} -\langle\!\langle g^{\mu\nu} T_{R\mu\nu}(x) \rangle\!\rangle\Big\vert_{odd} ={-  \frac 1{96\pi^2}}\epsilon_{\mu\nu\lambda\rho} \,\partial^\mu V^\nu(x) \partial^\lambda V^\rho(x)\label{tracedifferenceb}
\ee

\section{A heat-kernel derivation. The SDW method for Weyl fermions}

We would like now to compare the previous result with the one obtained with a Seeley-DeWitt method. In the sequel we apply this method to a right-handed Weyl fermion to compute the odd-parity trace anomaly. But, as we shall see, and as it should be expected at this point, this calculation is strictly connected to the calculation of the consistent chiral anomaly. It is in fact more convenient to carry out the latter first. 

\subsection{The consistent chiral anomaly via SDW}

The general method is the same as in section 3 but for the choice of the appropriate elliptic operator. To put the problem in a suitable framework we consider a Dirac fermion in a non-Abelian $(V,A)$ background, i.e. we use the covariant operator $i\slashed \nabla$, defined as follows 
\be
\nabla_\mu =\partial_\mu +\frac 12  \omega_\mu + \widehat {\EV}_\mu \label{nablamu}
\ee
where $\widehat\EV_\mu = V_\mu +\gamma_5 A_\mu$ and $V_\mu = V_\mu^a T^a, A_\mu =A_\mu ^a T^a$, and $T^a$ are anti-hermitean Lie algebra generators. Next we take the limit $V\to V/2, A\to V/2$. Then the relevant kinetic operator becomes
\be
{\slashed{\mathscr D}}=i \gamma^\mu {{\mathscr D}}_\mu, \quad\quad
{\mathscr D}_\mu= \partial_\mu + P_+ V_\mu, \quad\quad\overline{{\mathscr D}}_\mu= \partial_\mu + P_- V_\mu, \quad\quad P_{\pm}= \frac {1\pm \gamma_5}2\label{mathscrD}
\ee
As pointed out before, in order to apply the SDW method we need the square of this operator, which, however, as before, is not selfadjoint
\be
\left({\slashed{\mathscr D}}^2\right)^\dagger= \gamma_0 {\slashed{\mathscr D}}^2\gamma_0\label{D2dagger}
\ee
After a Wick rotation this relation becomes
\be
\left(\widetilde{\slashed{\mathscr D}}^2\right)^\dagger= \widetilde{\slashed{\mathscr D}}^2\label{widetildeDsquare}
\ee
Therefore we can use $\widetilde{\slashed{\mathscr D}}^2$. But, in practice, with the precautions explained in section 2.3, we will work with
\be
{\slashed{\mathscr D}}^2 &=& -\eta^{\mu\nu}\overline{{\mathscr D}}_\mu
{{\mathscr D}}_\nu- \Sigma^{\mu\nu} \left(\overline{{\mathscr D}}_\mu
{{\mathscr D}}_\nu-\overline{{\mathscr D}}_\nu
{{\mathscr D}}_\mu\right)\0\\
&=& -\left(\square + P_+ \partial\!\cdot\! V + V\!\cdot\! \partial\right) -\Sigma^{\mu\nu}\left(\partial_\mu V_\nu- \partial_\nu V_\mu + \gamma_5 \left( V_\nu\partial_\mu\  -V_\mu\partial_\nu\right)\right)\label{mathscrD2}
\ee

The consistent chiral anomaly is given by 
\be
{\cal A}=\!\! \int d^4x\left(  \lim_{x'\to x}\int_0^\infty ds \, \Tr \left( 2\gamma_5\, \rho \,\langle x,s|x',0\rangle\right) \right)=\!\! \int d^4x \left(\lim_{x'\to x}\int_0^\infty ds \, \Tr \left( \gamma_5 \,\rho\, \langle x|e^{i\EF s} |x'\rangle\right)\right)\label{calAtr}
\ee
where $\rho=\rho^a (x)T^a$ is the gauge parameter and $\EF=\widetilde{\slashed{\mathscr D}}^2$ and $\Tr$ includes all the traces. Expanding $\EF$ in powers of $is$ this becomes
 \be
{\cal A}= \frac i{16\pi^2} \int d^4x\Tr \left( \gamma_5 \,\rho\, [a_2(x)]\right),\quad\quad [a_2(x)] = \lim_{x'\to x} a_2(x,x')
\label{calAtr2}
\ee
In order to compute the coefficient $a_2$ in a flat background we have at our disposal the heat kernel equation
\be
i\frac{\partial}{\partial s} \langle x,s|x',0\rangle = - {{\slashed{\mathscr D}}^2}\langle x,s|x',0\rangle \label{heatkernelflat}
\ee
with
\be
\langle x,s|x',0\rangle =-\frac 1{(4\pi \alpha)^2 } e^{i\frac {(x-x')^2}{4 s}}\,\Phi(x,x';s)\label{xalphax'flat}
\ee
Replacing this into \eqref{heatkernelflat} and using \eqref{mathscrD2} we get the equation
\be
i\frac {\partial \Phi}{\partial s}\!\!&+&\!\!\square \Phi+V^\mu \partial_\mu \Phi + \frac i s (x\!-\!x')^\mu \partial_\mu \Phi+ \frac i{2s} (x\!-\!x')^\mu V_\mu \Phi +\frac is\gamma_5\Sigma^{\mu\nu} (x\!-\!x')_\mu V_\nu \Phi\0\\
&&-2 \gamma_5\Sigma^{\mu\nu}V_\mu \partial_\nu \Phi +P_+ \bigl( \partial\!\cdot\! V + \Sigma^{\mu\nu} (\partial_\mu V_\nu- \partial_\nu V_\mu)\bigr)\Phi=0 \label{heatkernelflat2}
\ee
Using the expansion 
\be
\Phi( x, x', s)
= \sum_{n=0}^\infty  a_n( x, x') (i s)^n
\label{Phiexp2}
\ee
we arrive at the recursion relation
\be
(n\!+\!1)a_{n+1}\!\!&+&\!\!\square a_n +V^\mu \partial_\mu a_n+  (x\!-\!x')^\mu \partial_\mu a_{n+1}+\frac 12 (x\!-\!x')^\mu V_\mu a_{n+1}+\gamma_5\Sigma^{\mu\nu} (x\!-\!x')_\mu V_\nu a_{n+1}\0\\
&&-2 \gamma_5\Sigma^{\mu\nu}V_\mu \partial_\nu a_n+P_+ \bigl( \partial\!\cdot\! V + \Sigma^{\mu\nu} (\partial_\mu V_\nu- \partial_\nu V_\mu)\bigr)a_n=0 \label{heatkernelflat2}
\ee
Setting $n=-1$ we get 
\be
(x\!-\!x')^\mu \partial_\mu a_0+\frac 12 (x\!-\!x')^\mu V_\mu a_0+\gamma_5\Sigma^{\mu\nu} (x\!-\!x')_\mu V_\nu a_0=0\label{n=-1}
\ee
Differentiating with respect to $x^\mu$ we obtain
\be
&&\partial_\mu a_0+\frac 12V_\mu a_0+\gamma_5\Sigma_\mu{}^{\nu} V_\nu a_0+
(x\!-\!x')^\lambda \partial_\mu\partial_\lambda a_0+\frac 12 (x\!-\!x')^\lambda\partial_\mu V_\lambda  a_0\0\\
&&\quad\quad+ \frac 12 (x\!-\!x')^\lambda V_\lambda\partial_\mu a_0+\gamma_5\Sigma^{\lambda\nu} (x\!-\!x')_\lambda \partial_\mu V_\nu a_0+\gamma_5\Sigma^{\lambda\nu} (x\!-\!x')_\lambda  V_\nu \partial_\mu a_0 =0\label{n=-1bis}
\ee
Starting from $[a_0](x)=1$ we compute the coincidence limit and obtain, for instance, $
[\partial_\mu a_0]= -\frac 12V_\mu -\gamma_5\Sigma_\mu{}^{\nu} V_\nu$.
Then, differentiating \eqref{n=-1bis} with respect to $x^\lambda$, contracting $\lambda$ with $\mu$ and taking the coincidence limit we get
\be
[\square a_0]= -\frac12 \partial\!\cdot\! V -\frac 38 V^2 - \Sigma^{\mu\nu} V_\mu V_\nu -\frac 12 \gamma_5 \Sigma^{\mu\nu} \left(\partial_\mu V_\nu -\partial_\nu V_\mu\right)\label{boxa0}
\ee
where we have used $\Sigma^{\mu\nu}\Sigma_\mu{}^\lambda = -\frac 58 \eta^{\nu\lambda} -\Sigma^{\nu\lambda}$. And so on.

Similarly, choosing now $n=0$ and proceeding the same way we get
\be
[a_1] &=& -\frac12 \partial\!\cdot\! V -\frac 38 V^2 + \Sigma^{\mu\nu} V_\mu V_\nu -\frac 12 \gamma_5 \Sigma^{\mu\nu} \left(\partial_\mu V_\nu -\partial_\nu V_\mu\right)+ P_+ \left( \partial\!\cdot\! V+ \gamma_5 \Sigma^{\mu\nu} \left(\partial_\mu V_\nu -\partial_\nu V_\mu\right)\right)\0\\
&=&  \frac 38 V^2 + \Sigma^{\mu\nu} V_\mu V_\nu+\frac 12 \left( \gamma_5\partial\!\cdot\! V+\Sigma^{\mu\nu} \left(\partial_\mu V_\nu -\partial_\nu V_\mu\right)\right)   \label{a14d}
\ee
and, with $n=1$,
\be
[a_2] = \frac 12 [\square a_1] +\frac 12 V^\lambda [\partial_\lambda a_1] +\gamma_5 \Sigma^{\lambda \nu} V_\nu[\partial_\lambda a_1] +\frac 12 P_+ \bigl(\partial\!\cdot\! V+\Sigma^{\mu\nu}C_{\mu\nu} [a_1]\bigr)\label{a24d}
\ee
where $C_{\mu\nu} = \partial_\mu V_\nu- \partial_\nu V_\mu$. We need to compute $ [\partial_\lambda a_1]$ and $ [\square a_1] $, which, in turn, requires the knowledge of $ [\square\square a_0] ,  [ \partial_\mu\square a_0]$ and $ [\partial_\lambda \partial_\nu a_0]$. All these quantities can be computed with a lenghty but straightforward procedure.

Replacing them into \eqref{a24d} we find a long expression, but only few terms can contribute to the odd parity part: they must be proportional to four gamma's, so it is not hard to figure out the candidates. For $[a_2]$ they are the following ones
\be
[a_2]&=& \frac16 V^\lambda V^\mu  [\partial_\lambda \partial_\mu a_0]+\frac 13 \Sigma^{\nu\rho}V_\nu V_\rho [a_1] +\frac 23 \Sigma^{\lambda\nu} \Sigma^{\mu\rho} V_\nu V_\rho[ \partial_\lambda \partial_\mu a_0] +\frac 13 \gamma_5  \Sigma^{\mu\nu} \{V^\lambda, V_\nu\}   [\partial_\lambda \partial_\mu a_0]\0\\
&&+\left( \frac14 +\frac 1{12} \gamma_5\right) \Sigma^{\mu\nu} C_{\mu\nu} [a_1]+\frac 16 \partial^\lambda V^\mu   [\partial_\lambda \partial_\mu a_0]+\frac 13 \gamma_5 \Sigma^{\mu\nu} \partial_\lambda V_\nu   [\partial_\lambda \partial_\mu a_0]\0\\
&&+ \frac 14 V^\lambda [\partial_\lambda \square a_0] +\frac 12 \gamma_5 \Sigma^{\mu\nu} V_\nu  [\partial_\lambda \square a_0] +\frac 1{12} \Sigma^{\mu\nu} C_{\mu\nu} [\square a_0]+\ldots \label{a24dext}
\ee
where ellipses denote terms that cannot contribute to the odd parity part.

Using the above expansions to evaluate \eqref{calAtr2} and
$\tr (\gamma_5 \gamma_\mu\gamma_\nu \gamma_\lambda \gamma_\rho)=-4i \epsilon_{\mu\nu\lambda\rho}$, we find that the term with 4 $V$'s vanish. The coefficients of the terms $\partial V V V, V\partial V V$ and $VV\partial V$ are all equal, and equal half the coefficient of the term $\partial V \partial V$. In conclusion we obtain the well-known expression
\be
{\cal A}= \frac i{16\pi^2} \int d^4x\Tr \left( \gamma_5 \,\rho\, [a_2(x)]\right)=- \frac 1{24\pi^2} \int d^4x\,\epsilon^{\mu\nu\lambda \rho} \, \tr\left( \partial_\mu \rho\left(  V_\nu \partial_\lambda V_\rho +\frac 12 V_\nu V_\lambda V_\rho\right)\right)\label{gaugeanom4d}
\ee
In the Abelian case we have $C_{\mu\nu}=V_{\mu\nu}$ and, of course, the second term in the RHS is absent. In view of the following subsection we notice that the result would be the same if, in the mid terms of \eqref{gaugeanom4d}, instead of $\gamma_5$ there were $2P_+$, for the difference in the two cases vanishes for parity odd terms.

{It should be stressed that in the above derivation we have dropped all the even parity terms. They are polynomials of canonical dimension 4 in $V$ and its derivatives, which satisfy the WZ consistency conditions. Since we know there do not exist even parity chiral anomalies, these terms can be canceled by adding suitable counterterms to the effective action. We dispense from that here.}

\subsection{The odd trace anomaly via SDW}

We use the previous calculation as a shortcut to compute the odd trace anomaly for a Weyl fermion coupled to a vector potential $V_\mu$. In this case we need to introduce an axial partner also for the metric, that is we couple the Weyl fermion to a MAT (metric-axial-tensor) background, \cite{BCDDGS}. The kinetic operator in this case is $i\slashed {\mathbb D}$ with
\be
{\mathbb D}_\mu =\widehat D_\mu +\frac 12 \widehat \Omega_\mu + V_\mu \label{nablamu}
\ee
 where $\widehat D_\mu$ is the covariant derivative  and  $\widehat \Omega_\mu= \Omega^{(1)}_\mu +\gamma_5 \Omega_\mu^{(2)}$ the spin connection with respect to the metric $\widehat g_{\mu\nu} = g_{\mu\nu} + \gamma_5 f_{\mu\nu}$. Here $V_\mu$ is an imaginary 
vector potential in agreement with the notation of section 2 and the previous subsection. In reality, for our present calculation, we will not need the full SDW derivation in a MAT background. A shortcut is as follows. The density of the anomaly we are after is $\rho\,\epsilon^{\mu\nu\lambda\rho} \partial_{\mu}V_\nu \partial_\lambda V_\rho$ which `occupies' the dimensionally available slots in 4d. The only change that MAT can (and does) bring about is the multiplication by the factor $\sqrt{\widehat g}\, \widehat\omega$, where
$ \widehat\omega= \omega+ \gamma_5 \eta$ is the extended Weyl parameter. Therefore, giving for granted a derivation analogous section 3 and 5.1, and using the $\zeta$-function regularization as in \cite{BCDGPS}, we can simply replace the symbols in the final formula. The variation of the effective action under an extended infinitesimal Weyl transformation
\be
\delta_{\widehat \omega}\widehat L  &=& -i \Tr \left( \widehat \omega \,
\zeta(x,0)\right)_c\label{anomhatomega}\\
&=&i\,\Tr \left(\frac {\sqrt{\widehat g}}{2 (4\pi)^2}\widehat \omega  \left(2[
a_2(x)] -2m^2
[a_1(x)] +m^4\right)\,\right)_c\0
\ee
where $\Tr$ denotes spacetime integration and gamma matrix trace and $[a_1], [a_2]$ are those of the previous subsection, and the subscript ${}_c$ denotes the chiral limit. The terms proportional to $m^2$ drop out because of the limit $m\to 0$. It remains for us to compute the chiral limit of $\sqrt{\widehat g}\,\widehat\omega$.
To this end let us recall
\be
{\sqrt{\widehat g} =\sqrt{\det (\widehat g)} = \sqrt{\det (g+\gamma_5 f)}}\label{volume1}
\ee
Using the formula $\det = e^{\tr \log}$ this gives
\be
 \tr\ln ( g+\gamma_5 f)=P_R\tr \ln (g+f)  +P_L\,  \tr\ln (g-f)\label{volume2}
\ee
For Weyl fermions we have to take the chiral limit. In the right-handed chiral limit we have not only $V_\mu\to V_\mu/2, A_\mu\to V_\mu/2$ but also $h_{\mu\nu} \to h_{\mu\nu}/2, f_{\mu\nu}\to h_{\mu\nu}/2$, and $\omega \to \omega/2, \eta\to \omega/2$. Therefore $\sqrt {\widehat  g}\,\widehat \omega \to  \frac 12 P_R \sqrt{g}$. An additional factor $\frac 12$ is due, as explained in section 2, to the fact that the kinetic operator we are working with transforms  under a Weyl transformation in a way complementary to \eqref{PsiWeyl}, i.e. with a parameter $-\frac 12 \omega$. Therefore we obtain
\be
\delta_\omega L= -\frac i{32\pi^2} \int d^4x\, \sqrt{g}\, \omega\,\tr \left( P_R [a_2](x)\right)\label{intLx}
\ee
Recalling now the remark in last paragraph of the previous section, $\tr \left( P_R [a_2](x)\right)$ produces precisely the density $\epsilon^{\mu\nu\lambda\rho} \partial_{\mu}V_\nu \partial_\lambda V_\rho$.  Therefore \eqref{intLx} yields
\be
{\cal A}_\omega = -\frac 1{96\pi^2} \int d^4x\,\sqrt{g}\, \omega\, \epsilon^{\mu\nu\lambda\rho} \partial_{\mu}V_\nu \partial_\lambda V_\rho,\label{Aomega''}
\ee
wich coincides with eq.\eqref{tracedifferenceb}.

In \cite{BSZ} this anomaly was tagged with the label  $^{(cs)}$, which stand for consistent\footnote{In fact in \cite{BSZ} only the second term in the RHS of \eqref{tracedifferenceb} was computed, which explains the different coefficients.}. The  origin of this label is better explained starting from an Abelian $V-A$ background. If we compute the trace anomaly due to such a background for a Dirac fermion we can take two limits: either the vector limit $A\to 0$ or the chiral limit $V\to V/2, A\to V/2$. In the two cases the trace anomaly has the same form but different coefficients, which we distinguish by labeling the first with $  ^{(cv)}$, the second with $^{(cs)}$.

\section{Anomalies and diffeomorphisms}

As already noticed, in order to validate the above results, it is necessary to verify the invariance under diffeomorphisms in presence of a gauge field $V$, for we cannot exclude a priori an anomaly of the type $\int d^4x \sqrt{g}\,\partial\!\cdot\! \xi\,\epsilon^{\mu\nu\lambda\rho} \partial_\mu V_\nu \partial_\lambda V_\rho$. This means that we need to analyze amplitudes $\langle \partial\!\cdot \! T\, J, J\rangle$, where $T$ and $J$ are the appropriate em tensor and currents. The proof in the case of (absence of) odd parity anomalies can be given in a general form, applicable also to a $V-A$ background. For this reason we consider amplitudes that involve $T_{\mu\nu}$ and $j_\mu$, but also $T_{5\mu\nu}$ and $j_{5\mu}$, where the last two are obtained by inserting $\gamma_5$ in the former.

\subsection{The amplitude  $\langle \partial\!\cdot\! T_5\, j\,j\rangle$}

Let us start from the amplitude  $\langle \partial\!\cdot\! T_5\, j\,j\rangle$, given by
\be
&&q^\mu \widetilde T^{(5VV)}_{\mu\nu\lambda\rho}(k_1,k_2)=\frac 14  \int
\frac{d^4p}{(2\pi)^4}\mathrm{tr}\left[\frac 1{\slashed{p}} \gamma_\lambda \frac 1 {\slashed{p}-\slashed{k}_1}\gamma_\rho\frac 1{\slashed{p}-\slashed{q}} \left(q\!\cdot\!(2p-q)\gamma_\nu +(2p-q)_\nu \slashed{q}\right)\gamma_5\right]\0\\
&=& \frac 14  \int
\frac{d^4pd^\delta\ell}{(2\pi)^{4+\delta}}\mathrm{tr}\left[\frac 1{\slashed{p}+\slashed{\ell}} \gamma_\lambda \frac 1 {\slashed{p}-\slashed{k}_1+\slashed{\ell}}\gamma_\rho\frac 1{\slashed{p}-\slashed{q}+\slashed{\ell}} \left(q\!\cdot\!(2p-q) \gamma_\nu+(2p-q)_\nu \slashed{q}\right)\gamma_5\right]\label{gauge&diff0}
\ee
We call $\widetilde A_{\nu\lambda\rho}(k_1,k_2)$ the piece proportional to $q\!\cdot\!(2p-q) \gamma_\nu$, and $\widetilde B_{\nu\lambda\rho}(k_1,k_2)$ the rest, and  write $q\!\cdot\!(2p-q) = 
p^2-\ell^2 -((p-q)^2-\ell^2)$. Then
\be
\widetilde A_{\nu\lambda\rho}(k_1,k_2)&=&  \frac 14  \int
\frac{d^4pd^\delta\ell}{(2\pi)^{4+\delta}}\biggl{\{}\frac{ -2^{2+\frac {\delta}2}i\epsilon_{\mu\nu\lambda\rho} \ell^2 (p+k_1-k_2)^\mu +\mathrm{tr}\left( (\slashed{p}+\slashed{k}_1)\gamma_\lambda \slashed{p} \gamma_\rho (\slashed{p}-\slashed{k}_2) \gamma_\nu \gamma_5\right) }{(p^2-\ell^2)((p-k_2)^2-\ell^2)}\0\\
&&+ \frac{ 2^{2+\frac {\delta}2}i\epsilon_{\mu\nu\lambda\rho} \ell^2 (p-k_2)^\mu -\mathrm{tr}\left(  \slashed{p} \gamma_\lambda (\slashed{p}-\slashed{k}_1)\gamma_\rho (\slashed{p}-\slashed{q}) \gamma_\nu \gamma_5\right) }{(p^2-\ell^2)((p-k_1)^2-\ell^2)}\biggr{\}}\label{gauge&diff1}
\ee
where the first line has been obtained with a shift $p\to p+k_1$. To this we have to add the cross contribution $\lambda\leftrightarrow, k_1\leftrightarrow k_2$. Now, since the trace of a matrix equals the trace of its transpose we get
\be
\mathrm{tr}\left(  \slashed{p} \gamma_\lambda (\slashed{p}-\slashed{k}_1)\gamma_\rho (\slashed{p}-\slashed{q}) \gamma_\nu \gamma_5\right) = \mathrm{tr}\left(  \gamma_5\gamma_\nu(\slashed{p}-\slashed{q}) \gamma_\rho(\slashed{p}-\slashed{k}_1)\gamma_\lambda\slashed{p} \right)\label{transpose}
\ee
Next, one can prove that, when integrated, 
\be
\frac{ \mathrm{tr}\left(  \gamma_5\gamma_\nu(\slashed{p}-\slashed{q}) \gamma_\rho(\slashed{p}-\slashed{k}_1)\gamma_\lambda\slashed{p} \right)} {(p^2-\ell^2)((p-k_1)^2-\ell^2)}= \frac{ \mathrm{tr}\left((\slashed{p}+\slashed{k}_1)\gamma_\lambda\slashed{p}\gamma_\rho(\slashed{p}-\slashed{q})\gamma_\nu \gamma_5\right)} {(p^2-\ell^2)((p-k_2)^2-\ell^2)}\0
\ee
This is obtained, once again,  with the exchanges $\lambda\leftrightarrow\rho, k_1\leftrightarrow k_2$, followed by the shift $p\to p+k_2$ and $p\to -p$. Therefore, since eventually we have to add the cross contribution, we see that the second term in the second line of \eqref{gauge&diff1} cancels the second in the first line. In a similar fashion one can prove that, upon integration, 
\be
\frac {(p-k_2)^\mu}{ (p^2-\ell^2)((p-k_1)^2-\ell^2)}= \frac {(p+k_1-k_2)^\mu} {(p^2-\ell^2)((p-k_2)^2-\ell^2)} \0
\ee
by shifting $p\to p+k_1$ and changing $p\to -p$, followed by the cross exchange  $\lambda\leftrightarrow\rho, k_1\leftrightarrow k_2$. It follows that the first term in the first line of \eqref{gauge&diff1} cancels the first term of the second line. Therefore
\be
\widetilde A_{\nu\lambda\rho}(k_1,k_2)+\widetilde A_{\nu\rho\lambda}(k_2,k_1) =0 \label{gauge&diff2}
\ee

Now let us consider $\widetilde B$:
\be
&&\!\!\!\!\!\!\!\widetilde B_{\nu\lambda\rho}(k_1,k_2)=  \frac 14  \int
\frac{d^4pd^\delta\ell}{(2\pi)^{4+\delta}}\mathrm{tr}\left[\frac 1{\slashed{p}+\slashed{\ell}} \gamma_\lambda \frac 1 {\slashed{p}-\slashed{k}_1+\slashed{\ell}}\gamma_\rho\frac 1{\slashed{p}-\slashed{q}+\slashed{\ell}}(2p-q)_\nu \slashed{q}\gamma_5\right]\0\\
&=&\frac 14  \int
\frac{d^4pd^\delta\ell}{(2\pi)^{4+\delta}}\frac{ -2^{2+\frac {\delta}2}i\epsilon_{\mu\sigma\lambda\rho} \ell^2 (p+k_1)^\mu q^\sigma +\mathrm{tr}\left( (\slashed{p}+\slashed{q})\gamma_\lambda   (\slashed{p}+\slashed{k}_2)\gamma_\rho \slashed{p}\slashed{q} \gamma_5\right) }{(p^2-\ell^2)((p+k_2)^2-\ell^2)((p+q)^2-\ell^2)} (2p+q)_\nu\label{gauge&diffB}
\ee
after a shift $p\to p+q$. Now using transposition inside the trace
\be
\mathrm{tr}\left( (\slashed{p}+\slashed{q})\gamma_\lambda   (\slashed{p}+\slashed{k}_2)\gamma_\rho \slashed{p}\slashed{q} \gamma_5\right) =
- \mathrm{tr}\left( \slashed{p}\gamma_\rho  (\slashed{p}+\slashed{k}_2) \gamma_\lambda(\slashed{p}+\slashed{q})\slashed{q} \gamma_5\right)\0
\ee
The cross contribution $\lambda\leftrightarrow\rho, k_1\leftrightarrow k_2$ is
\be
&&\!\!\!\!\!\!\!\widetilde B_{\nu\rho\lambda}(k_2,k_1)= \frac 18  \int
\frac{d^4pd^\delta\ell}{(2\pi)^{4+\delta}}\frac{ 2^{2+\frac {\delta}2}i\epsilon_{\mu\sigma\lambda\rho} \ell^2 (p+k_2)^\mu q^\sigma - \mathrm{tr}\left( \slashed{p}\gamma_\lambda  (\slashed{p}+\slashed{k}_1) \gamma_\rho(\slashed{p}+\slashed{q})\slashed{q} \gamma_5\right) }{(p^2-\ell^2)((p+k_1)^2-\ell^2)((p+q)^2-\ell^2)} (2p+q)_\nu\0\\
&=&\frac 14  \int
\frac{d^4pd^\delta\ell}{(2\pi)^{4+\delta}}\frac{2^{2+\frac {\delta}2} i\epsilon_{\mu\sigma\lambda\rho} \ell^2 (p-k_2)^\mu q^\sigma - \mathrm{tr}\left( \slashed{p}\gamma_\lambda  (\slashed{p}-\slashed{k}_1) \gamma_\rho(\slashed{p}+\slashed{q})\slashed{q} \gamma_5\right) }{(p^2-\ell^2)((p-k_1)^2-\ell^2)((p-q)^2-\ell^2)} (2p-q)_\nu\label{gauge&diffB1}
\ee
Therefore
\be
\widetilde B_{\nu\lambda\rho}(k_1,k_2)+\widetilde B_{\nu\rho\lambda}(k_2,k_1)=0\label{gauge&diffB2}
\ee
Therefore
\be
\langle \partial\!\cdot\! T_5\, j\,j\rangle=0\label{dT5jj}
\ee

{\bf Remark.} For later use it is important to remark that the amplitude \eqref{gauge&diff0} vanishes separately for the $\ell^2$-dependent and the $\ell^2$-independent parts.

\subsection{The amplitude  $\langle \partial\!\cdot\! T_R\, j_R\,j_R\rangle$}

The triangle contribution is
\be
&&\!\!\!\!\!\!\!\! q^\mu \widetilde T^{(RRR)}_{\mu\nu\lambda\rho}(k_1,k_2)=\frac 14\!\int\!\!\!
\frac{d^4p}{(2\pi)^4}\mathrm{tr}\left[\frac 1{\slashed{p}} \gamma_\lambda P_R\frac 1 {\slashed{p}-\slashed{k}_1}\gamma_\rho P_R\frac 1{\slashed{p}-\slashed{q}} \left(q\!\cdot\!(2p-q)\gamma_\nu +(2p-q)_\nu \slashed{q}\right)P_R\right]\label{gauge&diffR0}\\
&=&\!\!\! \frac 14  \int
\frac{d^4pd^\delta\ell}{(2\pi)^{4+\delta}}\mathrm{tr}\left[\frac {\slashed{p}}{p^2-\ell^2} \gamma_\lambda \frac  {\slashed{p}-\slashed{k}_1}{(p-k_1)^2-\ell^2}\gamma_\rho\frac {\slashed{p}-\slashed{q}}{(p-q)^2-\ell^2} \left(q\!\cdot\!(2p-q) \gamma_\nu+(2p-q)_\nu \slashed{q}\right)\frac{ 1+\gamma_5}2\right]\0
\ee
It is evident that the odd part is half the $\ell^2$ independent part of \eqref{gauge&diff0}, thus it vanishes.

\subsection{The amplitude  $\langle \partial\!\cdot\! T_5\, j_5\,j_5\rangle$}

The amplitude is
\be
&&q^\mu \widetilde T^{(555)}_{\mu\nu\lambda\rho}(k_1,k_2)=\frac 14  \int
\frac{d^4p}{(2\pi)^4}\mathrm{tr}\left[\frac 1{\slashed{p}} \gamma_\lambda\gamma_5\frac 1 {\slashed{p}-\slashed{k}_1}\gamma_\rho\gamma_5\frac 1{\slashed{p}-\slashed{q}} \left(q\!\cdot\!(2p-q)\gamma_\nu +(2p-q)_\nu \slashed{q}\right)\gamma_5\right]\0\\
&=& \frac 14  \int
\frac{d^4pd^\delta\ell}{(2\pi)^{4+\delta}}\mathrm{tr}\left[\frac 1{\slashed{p}+\slashed{\ell}} \gamma_\lambda\gamma_5 \frac 1 {\slashed{p}-\slashed{k}_1+\slashed{\ell}}\gamma_\rho\gamma_5\frac 1{\slashed{p}-\slashed{q}+\slashed{\ell}} \left(q\!\cdot\!(2p-q) \gamma_\nu+(2p-q)_\nu \slashed{q}\right)\gamma_5\right]\label{gauge&diff5550}
\ee
The $\ell^2$-independent part is the same as before, and vanishes. The $\ell^2$-dependent part is
\be
&&q^\mu \widetilde T^{(555)}_{\mu\nu\lambda\rho}(k_1,k_2)\label{gauge&diffT555}\\
&=&2^{\frac {\delta}2-2} i\int \frac{d^4pd^\delta\ell}{(2\pi)^{4+\delta}}\,\ell^2\, \frac {\epsilon_{\mu\nu\lambda\rho} (3p\!-\!2k_1\!-\!k_2)^\mu q\!\cdot\!(2p\!-\!q) + \epsilon_{\mu\sigma\lambda\rho}  (3p\!-\!2k_1\!-\!k_2)^\mu q^\sigma (2p\!-\!q)_\nu)}{(p^2-\ell^2)((p-k_1)^2-\ell^2)((p-q)^2-\ell^2)} \0
\ee
If we shift $p\to p+q$, change $p\to -p$ and make the exchange $\lambda\leftrightarrow\rho, k_1\leftrightarrow k_2$ we obtain the same expression with opposite sign. Therefore the amplitude  $\langle \partial\!\cdot\! T_5\, j_5\,j_5\rangle$ vanishes too.

\subsection{The amplitudes  $\langle \partial\!\cdot\! T\, j\,j_5\rangle$ and  $\langle \partial\!\cdot\! T\, j_5\,j\rangle$}

The amplitudes  $\langle \partial\!\cdot\! T\, j\,j_5\rangle$ and  $\langle \partial\!\cdot\! T\, j_5\,j\rangle$ are nonvanishing, but in fact, due to the bosonic symmetry we have to compute the average $\langle \partial\!\cdot\! T\, j\,j_5\rangle+\langle \partial\!\cdot\! T\, j_5\,j\rangle$, that is 
\be
&&  q^\mu  \left(\widetilde T^{(VV5)}_{\mu\nu\lambda\rho}(k_1,k_2)+\widetilde T^{(V5V)}_{\mu\nu\lambda\rho}(k_1,k_2)\right) \0\\
&=& \frac 1{4} \int
\frac{d^4pd^\delta\ell}{(2\pi)^{4+\delta}}\mathrm{tr}\biggl{[}\frac 1{\slashed{p}+\slashed{\ell}} \gamma_\lambda \frac 1 {\slashed{p}-\slashed{k}_1+\slashed{\ell}}\gamma_\rho\gamma_5\frac 1{\slashed{p}-\slashed{q}+\slashed{\ell}} \left(q\!\cdot\!(2p-q) \gamma_\nu+(2p-q)_\nu \slashed{q}\right) \0\\
&& +\frac 1{\slashed{p}+\slashed{\ell}} \gamma_\lambda\gamma_5\frac 1 {\slashed{p}-\slashed{k}_1+\slashed{\ell}}\gamma_\rho\frac 1{\slashed{p}-\slashed{q}+\slashed{\ell}} \left(q\!\cdot\!(2p-q) \gamma_\nu+(2p-q)_\nu \slashed{q}\right)\biggr{]}\0\\
&=&- 2^{\frac {\delta}2} i\int \frac{d^4p d^\delta\ell}{(2\pi)^{4+\delta}}\,\ell^2\, \frac {\epsilon_{\mu\nu\lambda\rho} (p-k_1)^\mu\,  q\!\cdot\!(2p\!-\!q) + \epsilon_{\mu\sigma\lambda\rho} (p-k_1)^\mu q^\sigma (2p-q)_\nu }{(p^2-\ell^2)((p-k_1)^2-\ell^2)((p-q)^2-\ell^2)}\label{gauge&diffVV50}
\ee
Now if we shift $p\to p+q$, change $p\to -p$ and make the exchange $\lambda\leftrightarrow\rho, k_1\leftrightarrow k_2$ we obtain the same expression with opposite sign. Therefore \eqref{gauge&diffVV50} vanishes.

\vskip 1cm

This is enough to conclude that the WI's for diffeomorphims concerning odd amplitudes are conserved. The odd parity 2pt correlators vanish, therefore the odd parity WI's reduce to the vanishing of the divergence of 3pt functions containing one em tensor insertion. These 3pt correlator divergences must vanish, up to anomalies. And this is what we have just proven.

In conclusion, diffeomorphisms are conserved as far as the gauge fields are concerned: no odd parity anomalies appear.

\section{Conclusion}

We can now draw some conclusions concerning the formula \eqref{Duff}. On the basis of the above results we have the following situation: the WI \eqref{tjjlevel2} and \eqref{WItjj} says that perturbative
$\langle\!\langle g^{\mu\nu} T_{\mu\nu}(x) \rangle\!\rangle$ is not anomalous, while
$ g^{\mu\nu}\langle\!\langle T_{\mu\nu}(x) \rangle\!\rangle $ is anomalous because of   \eqref{trueAomega1}.

In the odd parity case the situation is different. Since all odd two-current correlators vanish, the WI reduces to the vanishing of the three-point functions em-tensor-current-current, that is
$ \langle 0| {\cal T} T_{R\mu}^{\mu}(x) j_R^\lambda(y) j_R^\rho(z) |0\rangle $ in the right-handed fermion case, and $ \langle 0| {\cal T} T_{\mu}^{\mu}(x) j_5^\lambda(y) j^\rho(z) |0\rangle$,  
$\langle 0| {\cal T} T_{5\mu}^{\mu}(x) j^\lambda(y) j^\rho(z) |0\rangle$  and
 $\langle 0| {\cal T} T_{5\mu}^{\mu}(x) j_5^\lambda(y) j_5^\rho(z) |0\rangle$ in the V-A case.
In the right-handed fermion case we have seen that both $ g^{\mu\nu}\langle\!\langle T_{R\mu\nu}(x) \rangle\!\rangle$  and $  \langle\!\langle T^\mu_{R\mu}(x) \rangle\!\rangle $ are anomalous with different anomalies (the same is true for the $\langle 0| {\cal T} T_{5\mu}^{\mu}(x) j^\lambda(y) j^\rho(z) |0\rangle$ case, which has not been reported here). 
To complete the panorama, we should add that, in the gravitational trace anomaly case for a right-handed Weyl fermion, the  $  \eta^{\mu\nu}\langle 0| T_{R\mu\nu}(x) T_{R\lambda \rho}(y) T_{R\sigma\tau}(z)|0\rangle $  vanishes, while  $ \langle 0| T_{R\mu}^\mu(x) T_{R\lambda \rho}(y) T_{R\sigma\tau}(z)|0\rangle $ is anomalous, \cite{BGL,BCDDGS,BCDGPS}. 

We would like now to spend a few words to interpret these results.

\subsection{A discussion about $g^{\mu\nu}\langle\!\langle T_{\mu\nu}(x) \rangle\!\rangle -\langle\!\langle g^{\mu\nu} T_{\mu\nu}(x) \rangle\!\rangle$}

The gauge and diffeomorphism anomalies are violations of the
classical conservation laws $\partial^\mu j_\mu(x)= 0$ and $\nabla^\mu T_{\mu\nu}(x)=0$, respectively. The trace anomaly is a violation of the classical tracelessness condition $T_\mu^\mu(x)=0$. The basic point here is that these equations are all valid on-shell, while off-shell they do not hold in general (except possibly in dimension 2). Another important point to be kept in mind is that, in terms of representations of the Lorentz group,  $T_{\mu\nu}(x)$ is a reducible tensor of which the trace $T_\mu^\mu(x)$ is an irreducible component. In the  expression of the effective action the latter is coupled to the field $h(x)=h_\mu^\mu(x)$. The amplitude $\langle 0|{\cal T} T_\mu^\mu (x)\Phi(y) \Psi(z)|0\rangle$, where $\Phi$ and $\Psi$ are generic fields, is an irreducible component of  $\langle 0|{\cal T} T_{\mu\nu}(x)\Phi(y) \Psi(z)|0\rangle$.

We have seen several examples where, when calculated with Feynman diagrams, the amplitudes  $\langle 0|{\cal T} T_\mu^\mu (x)\Phi(y) \Psi(z) |0\rangle$ and $\eta^{\mu\nu}\langle 0|{\cal T}_{\mu\nu}\Phi(y) \Psi(z)|0\rangle$ are generally different.  In the face of it, one possible attitude is to declare that the true value of the amplitude is given by the latter and ignore the former, considering their difference to be an oddity of the regularization. After all the (semi-classical, i.e. without anomaly) conformal WI  is satisfied also in this way. However there is a difficulty on the way of such a cavalier solution. On the one hand we have seen that the definition \eqref{Duff} for the trace anomaly coincides with the nonperturbative way of defining the trace anomaly. On the other hand we have seen on several examples, see also \cite{BDL}, that  the amplitudes of the type $\langle 0|{\cal T} T_\mu^\mu (x) j_{\lambda}(y) j_{\rho}(z)|0\rangle$ are related to the Adler-Bell-Jackiw anomalies, and, moreover, it is possible to prove that the amplitude $\langle 0|{\cal T} T_\mu^\mu (x) T_{\lambda\rho}(y) T_{\sigma\tau}(z)|0\rangle$ are rigidly related to the Kimura-Delbourgo-Salam anomaly, \cite{Kimura,DS}.
Therefore if we decide to ignore  $\langle 0|{\cal T} T_\mu^\mu (x) j_{\lambda}(y) j_{\rho}(z)|0\rangle$ and $\langle 0|{\cal T} T_\mu^\mu (x) T_{\lambda\rho}(y) T_{\sigma\tau}(z)|0\rangle$, the calculation of KDS and ABJ anomalies becomes problematic, to say the least\footnote{{For the connection between chiral trace and ABJ anomaly, see \cite{BSZ}; for the connection between chiral trace and Kimura-Delbourgo-Salam anomaly, see for instance ref.\cite{BDL}. In both cases the connection is visible at the level of the corresponding lowest order Feynman diagrams. But also in the non-perturbative approach of section 5, this link is clear. It should be added that this connection does not hold for even trace anomalies, and, in any case, although a link is undeniable, it certainly deserves a deeper analysis.}}.That is unacceptable.

Therefore we have to live with the trace anomaly (perturbatively) defined by the difference
\be
\eta^{\mu\nu}\langle 0|{\cal T} T_{\mu\nu}(x)\Phi(y) \Psi(z)  |0\rangle-\langle 0|{\cal T} T_\mu^\mu (x) \Phi(y) \Psi(z)  |0\rangle\label{difference}
\ee  
This difference means in particular that the (regularized) effective action has discontinuities: differentiating it with respect to $h_{\mu\nu}$ and then saturating the result with $\eta_{\mu\nu}$ is not the same as differentiating it with respect to $h(x)=h_\mu^\mu(x)$, which is the conjugate source of $T_\mu^\mu(x)$. Like in many other situations in quantum theories we are not allowed to make a choice such as ignoring the second term in \eqref{difference}. We have to let the theory speak and keep all the information provided by it and, eventually, interpret it.

A model of the situation we are facing is given by the formula
\begin{equation}
\square \frac 1{ x^2}\sim \delta^{(4)}\left(x\right).\label{squarelog}
\end{equation}
valid in distribution theory in a Euclidean  4d space. The derivatives $\partial_\mu$ of $\frac 1{x^2}$ or $\frac{x^\nu}{x^4}$ are  well defined for $x^\mu\neq 0$ where $\square \frac 1{x^2}=0$ , but they are ill-defined at $x^\mu=0$. On the other hand the derivation with respect to $r=\sqrt{x^2}$ makes sense even at $r=0$, and gives rise to the formula \eqref{squarelog}. 
In our case the analog of $x^\mu$ is $h_{\mu\nu}$, the analog of $r$ is $h_\mu^\mu$ and the analog of $x=0$ is the classical on-shell condition $T_\mu^\mu=0$ (for a closer analogy one should actually consider $\square\frac 1{ (x^2+a^2)^2}$ which becomes a delta function when $a\to 0$).
We are therefore forced to take into account this discontinuity of the effective action.
Indeed, in the perturbative approach, we are obliged to refine the naive definition of the trace anomaly as follows: 
\be 
g^{\mu\nu}(x)\langle\!\langle T_{\mu\nu}(x)\rangle\!\rangle-  \langle\!\langle g^{\mu\nu}(x) T_{\mu\nu}(x)\rangle\!\rangle=- {T}[g](x).\label{1pttraceanom2}
\ee
After verifying that this definition works properly, one may ask what its physical meaning is.
It is clear that the reason for taking the difference in the LHS of \eqref{1pttraceanom2} is that two correlators may in general contain extra terms which have nothing to do with the anomaly. These terms are
\begin{itemize}
\item possible soft terms that classically violate conformal invariance;
\item the term $i\eta^{\mu\nu} \bar \psi \slashed {\partial} \psi$ in the modified definition of the e.m. tensor;
\item the semilocal terms in the conformal WI;
\item possible off-shell contributions to the anomaly: contrary to the example above where $\square$ applied to the argument yields 0 for $x\neq 0$, the derivative with respect to $h_{\mu\nu}$ contracted with $\eta^{\mu\nu}$, or the derivative with respect to $h_\mu^\mu$, do not automatically vanish off-shell. In fact the operator $T_\mu^\mu$ identically vanish on shell, therefore its contribution can only be off-shell. This means that in formula \eqref{1pttraceanom2}  the off-shell contributions to the anomaly are subtracted away. In other words the trace anomaly \eqref{1pttraceanom2} receives only on-shell contributions.
\end{itemize}
All these terms cancel out in \eqref{1pttraceanom2}.

Let us expand a bit on the last point. First of all let us notice that 
\be
T_\mu^\mu(x)\sim \bar\psi \stackrel {\leftrightarrow}{\slashed\nabla} \psi\label{Tmumu}
\ee
Therefore $\langle 0|{\cal T} T_\mu^\mu (x) \Phi(y) \Phi(z)|0\rangle$ is proportional to the LHS of the equation of motion.  So it represents a contribution to the quantum object $\langle\!\langle \bar\psi \stackrel {\leftrightarrow}{\slashed\nabla} \psi \rangle\!\rangle$ off-shell (because on-shell it vanishes)

On the other hand $\eta^{\mu\nu}\langle 0|{\cal T} T_{\mu\nu}(x) \Phi(y) \Phi(z)|0\rangle$ represents the contribution coming from the differentiation of the effective action with respect to $h^{\mu\nu}(x)$, which we know does not coincide with the differentiation with respect to $h(x)$ (there is a discontinuity). Therefore we interpret it as the on-shell plus off-shell contribution  to $\langle\!\langle \bar\psi \stackrel {\leftrightarrow}{\slashed\nabla} \psi \rangle\!\rangle$. The difference \eqref{difference} measures the one-loop violation to the equation of motion.  It represents so to speak the quantization of 0, a genuine quantum effect, and supports formula \eqref{Duff}.  We call this violation {\it the trace anomaly.}

 This has to be compared (and contrasted) with the other anomalies (gauge and diff). For instance, in the case of gauge anomalies we have, similarly to the trace case,
\be
\partial_\mu j^\mu (x) \sim\partial_\mu \left(\bar \psi \gamma^\mu \psi\right)\label{partialmujmu}
\ee 
which is proportional to the LHS of the Dirac eoms. However in this case the amplitude 
$\langle0|{\cal T}\partial^\mu j_\mu(x) j_\nu(y) j_\lambda(z)|0\rangle$ gives the same result
as $\partial_x^\mu\langle0|{\cal T} j_\mu(x) j_\nu(y) j_\lambda(z)|0\rangle$. Which is not surprising because $\langle0|{\cal T}\partial^\mu j_\mu(x) j_\nu(y) j_\lambda(z)|0\rangle$ is not an amplitude independent of $\langle0|{\cal T}  j_\mu(x) j_\nu(y) j_\lambda(z)|0\rangle$. This is reflected in the fact that in the effective action $\langle0|{\cal T}\partial^\mu j_\mu(x) j_\nu(y) j_\lambda(z)|0\rangle$ is not coupled to an independent source field. There is no way to disentangle the on-shell from the off-shell part, if any.
Therefore we simply set
\be
\partial_\mu \langle\!\langle j^\mu(x) \rangle\!\rangle = \langle\!\langle\partial_\mu j^\mu(x) \rangle\!\rangle= {\rm anomaly}\label{partialjpartialj}
\ee

To conclude let us make a comment on the definition \eqref{Duff}. As we said at the beginning, it has a clear meaning in a perturbative framework. It is not applicable to a non-pertrubative approach, like the SDW one, in which case the definition of trace anomaly is simply the response of the effective action to a Weyl transformation. Therefore the natural question is: what is the meaning of the two terms in the LHS of \eqref{Duff} in relation to the non-perturbative case? {An explanation has been suggested in the introduction. The perturbative approach is based on the lowest order of the perturbative cohomology, which is a much looser mathematical structure than the full BRST cohomology (whose non-trivial cocycles are very limited in number). The definition \eqref{Duff} is taylored to channel the lowest order perturbative results in the direction of a coincidence with the non-perturbative approaches.}{In a more forbished language one could say the each term of \eqref{Duff} is unstable in terms of perturbative cohomology, while their difference is stable.}

\vskip 1cm

{\bf Acknowledgements}. I would like to thank Roberto Soldati for several stimulating and clarifying  discussions and Adam Schwimmer for a useful exchange of messages.
\vskip 1cm

{\small

\appendix

\section{Appendix A. Even gauge current correlators}

In this Appendix we show that in a theory of Dirac or Weyl fermions 
the integrated anomaly $\int d^4x \sqrt{g}\, \alpha(x)\,  F_{\mu\nu}(x) F^{\mu\nu}(x)$ cannot appear in the divergence of a gauge current, i.e. from regularizing an amplitude $\langle \partial \!\cdot \! J\, J_\lambda\,J_\rho\rangle$ (with an even number of $j_5$). We recall that, in order to reproduce the anomaly proportional to  $\int d^4x \sqrt{g}\, \alpha(x)\,  F_{\mu\nu}(x) F^{\mu\nu}(x)$, the Fourier transform of the amplitude must contain a local term of the form $\sim \left(k_{1\rho} k_{2\lambda}- \eta_{\lambda\rho} k_1 \!\cdot\! k_2\right)$. 

\subsection{A preliminary calculation}

As a starting  calculation we want to prove that the even triangle diagram contribution
\begin{eqnarray}
q^\mu \widetilde T_{\mu\lambda\rho}^{(even1)}(k_1,k_2) = \int
\frac{d^4p d^\delta\ell}{(2\pi)^{4+\delta}}\mathrm{tr}\left\{\frac{\slashed{p}}{{p}^2-{\ell}^2}     
\gamma_\lambda\frac{\slashed{p}
-\slashed{k}_1}{(p-k_1)^2 -\ell^2} \gamma_{\rho}
\frac{\slashed{p}-\slashed{q}}{(p-q)^2-\ell^2}
\slashed{q} \right\}\0\\
\label{triangledimeven}
\end{eqnarray}
vanishes. Using $(\slashed{p}- \slashed{q}) \slashed{q}\slashed{p}= (2p\!\cdot\!q-q^2) \slashed{p}- p^2 \slashed{q}$ , the integrand becomes 
\be
q^\mu \widetilde T_{\mu\lambda\rho}^{(even1)}(k_1,k_2) = \int
\frac{d^4p d^\delta\ell}{(2\pi)^{4+\delta}}\frac {(2p\!\cdot\!q-q^2)\mathrm{tr}\left(\slashed{p}\gamma_\lambda (\slashed{p}-\slashed{k_1})\gamma_\rho\right)- p^2 \mathrm{tr}\left(\slashed{q}\gamma_\lambda (\slashed{p}-\slashed{k_1})\gamma_\rho\right)}
{(p^2 -\ell^2) ((p-k_1)^2 -\ell^2)((p-q)^2-\ell^2)}\label{triangledimeven1}
\ee
Evaluating traces:
\be
&=& 2^{2+\frac {\delta}2}\int \frac{d^4p d^\delta\ell}{(2\pi)^{4+\delta}}\Bigl{[}-(p-q)^2 \bigl(p_\lambda (p-k_1)_\rho -\eta_{\lambda\rho}p\!\cdot\!(p-k_1) + p_\rho (p-k_1)_\lambda\bigr)\0\\
&&+p^2 \bigl( (p-q)_\lambda (p-k_1)_\rho -\eta_{\lambda\rho}(p-q)\!\cdot\! (p-k_1)+  (p-q)_\rho (p-k_1)_\lambda\bigr)\Bigr{]}\0\\
&&\times\, \frac 1{(p^2 -\ell^2) ((p-k_1)^2 -\ell^2)((p-q)^2-\ell^2)}\label{triangledimeven2}
\ee
To this we have to add the cross contribution $k_1\leftrightarrow k_2, \lambda\leftrightarrow \rho$ 
\begin{eqnarray}
q^\mu \widetilde T_{\mu\rho\lambda}^{(even2)}(k_2,k_1)& =&2^{2+\frac {\delta}2}\int \frac{d^4p d^\delta\ell}{(2\pi)^{4+\delta}}\Bigl{[}-(p-q)^2 \bigl(p_\lambda (p-k_2)_\rho -\eta_{\lambda\rho}p\!\cdot\!(p-k_2) + p_\rho (p-k_2)_\lambda\bigr)\0\\
&&+p^2 \bigl( (p-q)_\lambda (p-k_2)_\rho -\eta_{\lambda\rho}(p-q)\!\cdot\! (p-k_2)+  (p-q)_\rho (p-k_2)_\lambda\bigr)\Bigr{]} \0\\
&&\times \frac 1{(p^2 -\ell^2) ((p-k_2)^2 -\ell^2)((p-q)^2-\ell^2)}\label{triangledimeven3}
\ee
Now shift $p\to p+q$ and change $p\to -p$, then \eqref{triangledimeven3} becomes
\be
&&q^\mu \widetilde T_{\mu\rho\lambda}^{(even2)}(k_2,k_1) =2^{2+\frac {\delta}2}\int \frac{d^4p d^\delta\ell}{(2\pi)^{4+\delta}}\Bigl{[}-p^2 \bigl( (p-q)_\lambda (p-k_1)_\rho -\eta_{\lambda\rho}(p-q)\!\cdot\! (p-k_1)\0\\
&&\quad\quad+  (p-q)_\rho (p-k_1)_\lambda\bigr)+(p-q)^2 \bigl(p_\lambda (p-k_1)_\rho -\eta_{\lambda\rho}p\!\cdot\!(p-k_1) + p_\rho (p-k_1)_\lambda\bigr)\Bigr{]} \0\\
&&\quad\quad\times \frac 1{(p^2 -\ell^2) ((p-k_1)^2 -\ell^2)((p-q)^2-\ell^2)}\label{triangledimeven4}
\ee
which is the opposite of  \eqref{triangledimeven2}. Therefore
\be
q^\mu \widetilde T_{\mu\lambda\rho}^{(even1)}(k_1,k_2)+q^\mu \widetilde T_{\mu\rho\lambda}^{(even2)}(k_2,k_1)=0\label{triangledimeven5}
\ee

Now we can look at the various cases

\subsection{Even part of $\langle \partial\!\cdot \! j_R\, j_R\, j_R\rangle$}

The three point function  $\langle \partial\!\cdot \! j_R\, j_R\, j_R\rangle$ is
\begin{eqnarray}
&&q^\mu \widetilde T_{\mu\lambda\rho}^{(R)}(k_1,k_2) = \int
\frac{d^4p d^\delta\ell}{(2\pi)^{4+\delta}}\mathrm{tr}\left\{\frac{1}{\slashed{p}+\slashed{\ell}}     
\frac{1-\gamma_5}{2}\gamma_\lambda\frac{1}{\slashed{p}+\slashed{\ell}
-\slashed{k}_1}\frac{1-\gamma_5}{2} \gamma_{\rho}
\frac{1}{\slashed{p}+\slashed{\ell}-\slashed{q}
}\frac{1-\gamma_5}{2}\slashed{q} \right\}\0\\
&&= \int
\frac{d^4pd^\delta\ell}{(2\pi)^{4+\delta}}\mathrm{tr}\left\{\frac{\slashed{p} }{{p}^2-{\ell}^2}     
 \gamma_\lambda\frac{\slashed{p} 
-\slashed{k}_1}{(p-k_1)^2 -\ell^2} \gamma_{\rho}
\frac{\slashed{p} -\slashed{q}}{(p-q)^2-\ell^2}
\frac{1-\gamma_5}{2}\slashed{q} \right\}\0\\
&&\equiv \widetilde F^{(R)}_{\lambda\rho}(k_1,k_2,\delta)\label{triangledim2}
\end{eqnarray}
The even part thereof is 1/2 of \eqref{triangledimeven}, therefore it vanishes.
\vskip 1cm
In view of an application to the $V-A$ system, let us consider other even 3pt correlator.

\subsection{Even part of $\langle \partial\!\cdot \! j\, j\, j\rangle$}

Let us consider the correlator
\begin{eqnarray}
q^\mu \widetilde T_{\mu\lambda\rho}^{(VVV)}(k_1,k_2) &=& \int
\frac{d^4p }{(2\pi)^{4}}\mathrm{tr}\left\{\frac 1{\slashed{p}}     
\gamma_\lambda\frac 1{\slashed{p}
-\slashed{k}_1}\gamma_{\rho}
\frac 1{\slashed{p}-\slashed{q}}
\slashed{q} \right\}\0\\
& =& \int
\frac{d^4p d^\delta\ell}{(2\pi)^{4+\delta}}\mathrm{tr}\Big\{\frac{\slashed{p}+\slashed{\ell}}{{p}^2-{\ell}^2}     
\gamma_\lambda\frac{\slashed{p}-\slashed{k}_1+\slashed{\ell}}{(p-k_1)^2 -\ell^2} \gamma_{\rho}
\frac{\slashed{p}-\slashed{q}+\slashed{\ell}}{(p-q)^2-\ell^2}\slashed{q} \Big\}
\label{triangleVVV}
\end{eqnarray}
Using \eqref{triangledimeven5}, this reduces to
\begin{eqnarray}
q^\mu \widetilde T_{\mu\lambda\rho}^{(VVV)}(k_1,k_2) &=&\int\frac{d^4p d^\delta\ell}{(2\pi)^{4+\delta}} \,\ell^2\,\frac{ \mathrm{tr}\left[ \gamma_\lambda \gamma_\rho(\slashed{p}-\slashed{q}) \slashed{q}+\gamma_\lambda (\slashed{p}-\slashed{k}_1) \gamma_\rho \slashed{q}+\slashed{p} \gamma_\lambda \gamma_\rho\slashed{q}\right]}{(p^2 -\ell^2) ((p-k_1)^2 -\ell^2)((p-q)^2-\ell^2)}\0\\
&=&4 \int\frac{d^4p d^\delta\ell}{(2\pi)^{4+\delta}}\, \ell^2\,\frac{\eta_{\lambda\rho} (p-k_2)\!\cdot\! q+q_\lambda (p-2k_1-k_2)_\rho +q_\rho (p+k_2)_\lambda}{(p^2 -\ell^2) ((p-k_1)^2 -\ell^2)((p-q)^2-\ell^2)}\label{triangleVVV1}
\ee
Introducing $x,y$ Feynman parameters and shifting $p\to p+(x+y)k_1+yk_2$, and integrating over $p$ and $\ell$, one gets
\be
q^\mu \widetilde T_{\mu\lambda\rho}^{(VVV)}(k_1,k_2) &=&-\frac{8i }{(4\pi)^2}\int_0^1 dx \int_0^{1-x}dy \Big[\eta_{\lambda\rho} ((x+y)k_1 +(y-1)k_2 )\!\cdot\! q\0\\
&& +q_\lambda ((x+y-2)k_1 +(y-1)k_2 )_\rho +q_\rho ((x+y)k_1 +(y+1)k_2 )_\lambda\Big]\0\\
&=& \frac i{12\pi^2} \Big[\eta_{\lambda\rho}(k_1-k_2)\!\cdot\! q -q_\lambda (2k_1+k_2)_\rho +  q_\rho (k_1+2k_2)_\lambda\Big]\label{triangleVVV2}
\ee
Adding the cross term ($k_1\leftrightarrow k_2$, $\lambda \leftrightarrow \rho$) one gets 0.

\subsection{Even part of $\langle \partial\!\cdot \! j\, j_5\, j_5\rangle$}

 Let us write down the triangle contribution to  $\langle \partial\!\cdot \! j\, j_5\, j_5\rangle$. It is
\begin{eqnarray}
q^\mu \widetilde T_{\mu\lambda\rho}^{(VAA)}(k_1,k_2) &=& \int
\frac{d^4p }{(2\pi)^{4}}\mathrm{tr}\left\{\frac 1{\slashed{p}}     
\gamma_\lambda\gamma_5\frac 1{\slashed{p}
-\slashed{k}_1}\gamma_{\rho}\gamma_5
\frac 1{\slashed{p}-\slashed{q}}
\slashed{q} \right\}\0\\
& =& \int
\frac{d^4p d^\delta\ell}{(2\pi)^{4+\delta}}\mathrm{tr}\Big\{\frac{\slashed{p}+\slashed{\ell}}{{p}^2-{\ell}^2}     
\gamma_\lambda\gamma_5\frac{\slashed{p}-\slashed{k}_1+\slashed{\ell}}{(p-k_1)^2 -\ell^2} \gamma_{\rho}\gamma_5
\frac{\slashed{p}-\slashed{q}+\slashed{\ell}}{(p-q)^2-\ell^2}\slashed{q} \Big\}
\label{triangleVAA}
\end{eqnarray}
Using \eqref{triangledimeven5}, this reduces to
\begin{eqnarray}
q^\mu \widetilde T_{\mu\lambda\rho}^{(VAA)}(k_1,k_2) &=&\int\frac{d^4p d^\delta\ell}{(2\pi)^{4+\delta}} \,\ell^2\,\frac{ \mathrm{tr}\left[ -\gamma_\lambda \gamma_\rho(\slashed{p}-\slashed{q}) \slashed{q}+\gamma_\lambda (\slashed{q}-\slashed{k}_1 \gamma_\rho \slashed{q}-\slashed{p} \gamma_\lambda \gamma_\rho\slashed{q}\right]}{(p^2 -\ell^2) ((p-k_1)^2 -\ell^2)((p-q)^2-\ell^2)}\label{triangleVAA1}\\
&=&4 \int\frac{d^4p d^\delta\ell}{(2\pi)^{4+\delta}}\, \ell^2\,\frac{-\eta_{\lambda\rho} (3p-2k_1-k_2)\!\cdot\! q+q_\lambda (p+k_2)_\rho +q_\rho (p-2k_1-k_2)_\lambda}{(p^2 -\ell^2) ((p-k_1)^2 -\ell^2)((p-q)^2-\ell^2)}\0
\ee
Introducing $x,y$ Feynman parameters and shifting $p\to p+(x+y)k_1+yk_2$, and integrating over $p$ and $\ell$, one gets
\be
q^\mu \widetilde T_{\mu\lambda\rho}^{(VAA)}(k_1,k_2) &=&-\frac{8i}{(4\pi)^2} \int_0^1 dx \int_0^{1-x}dy \Big[\eta_{\lambda\rho} ((3x+3y-2)k_1 +(3y-1)k_2 )\!\cdot\! q\0\\
&& +q_\lambda ((x+y)k_1 +(y+1)k_2 )_\rho +q_\rho ((x+y-2)k_1 +(y-1)k_2 )_\lambda\Big]\0\\
&=& \frac i{12\pi^2} \Big[-k_{1\lambda} k_{1\rho} +  k_{2\lambda} k_{2\rho}\Big]\label{triangleVAA2}
\ee
Adding the cross term ($k_1\leftrightarrow k_2$, $\lambda \leftrightarrow \rho$) one gets 0.

\subsection{Even part of $\langle \partial\!\cdot \! j_5\, j\, j_5\rangle$ and $\langle \partial\!\cdot \! j_5\, j_5\, j\rangle$}

Let us consider next the triangle contribution to $\langle \partial\!\cdot \! j_5\, j\, j_5\rangle$. It is
\begin{eqnarray}
q^\mu \widetilde T_{\mu\lambda\rho}^{(AVA)}(k_1,k_2) &=& \int
\frac{d^4p }{(2\pi)^{4}}\mathrm{tr}\left\{\frac 1{\slashed{p}}     
\gamma_\lambda\gamma_5\frac 1{\slashed{p}
-\slashed{k}_1}\gamma_{\rho}
\frac 1{\slashed{p}-\slashed{q}}\gamma_5
\slashed{q} \right\}\0\\
& =& \int
\frac{d^4p d^\delta\ell}{(2\pi)^{4+\delta}}\mathrm{tr}\Big\{\frac{\slashed{p}+\slashed{\ell}}{{p}^2-{\ell}^2}     
\gamma_\lambda\gamma_5\frac{\slashed{p}-\slashed{k}_1+\slashed{\ell}}{(p-k_1)^2 -\ell^2} \gamma_{\rho}
\frac{\slashed{p}-\slashed{q}+\slashed{\ell}}{(p-q)^2-\ell^2}\gamma_5\slashed{q} \Big\}
\label{triangleAVA}
\end{eqnarray}
Using \eqref{triangledimeven5}, this reduces to
\begin{eqnarray}
q^\mu \widetilde T_{\mu\lambda\rho}^{(AVA)}(k_1,k_2) &=&\int\frac{d^4p d^\delta\ell}{(2\pi)^{4+\delta}} \,\ell^2\,\frac{ \mathrm{tr}\left[ -\gamma_\lambda \gamma_\rho(\slashed{p}-\slashed{q}) \slashed{q}-\gamma_\lambda (\slashed{q}-\slashed{k}_1 \gamma_\rho \slashed{q}+\slashed{p} \gamma_\lambda \gamma_\rho\slashed{q}\right]}{(p^2 -\ell^2) ((p-k_1)^2 -\ell^2)((p-q)^2-\ell^2)}\label{triangleAVA1}\\
&=&4 \int\frac{d^4p d^\delta\ell}{(2\pi)^{4+\delta}}\, \ell^2\,\frac{\eta_{\lambda\rho} (p+k_2)\!\cdot\! q-q_\lambda (3p-2k_1-k_2)_\rho +q_\rho (p-k_2)_\lambda}{(p^2 -\ell^2) ((p-k_1)^2 -\ell^2)((p-q)^2-\ell^2)}\0
\ee
Introducing $x,y$ Feynman parameters and shifting $p\to p+(x+y)k_1+yk_2$, and integrating over $p$ and $\ell$, one gets
\be
q^\mu \widetilde T_{\mu\lambda\rho}^{(AVA)}(k_1,k_2) &=&-\frac {4i}{(4\pi)^2} \int_0^1 dx \int_0^{1-x}dy \Big[-\eta_{\lambda\rho} ((x+y)k_1 +(y+1)k_2 )\!\cdot\! q\0\\
&& +q_\lambda ((3x+3y-2)k_1 +(3y-1)k_2 )_\rho +q_\rho ((x+y)k_1 +(y-1)k_2 )_\lambda\Big]\0\\
&=& \frac i{4\pi^2} \Big[-\frac13\eta_{\lambda\rho}(k_1+2k_2)\!\cdot\!q  +\frac 13 q_\rho(k_1-k_2)_\lambda\Big]\label{triangleAVA2}
\ee
Adding the cross term one gets
\be
q^\mu \widetilde T_{\mu\lambda\rho}^{(AVA)}(k_1,k_2)+q^\mu \widetilde T_{\mu\rho\lambda}^{(AVA)}(k_2,k_1)= \frac i{4\pi^2} \Big[-\eta_{\lambda\rho}q^2  +\frac 23 (k_{1\lambda} k_{2\rho} -k_{2\lambda} k_{1\rho}\Big]\label{triangleAVA3}
\ee

The even part of $\langle \partial\!\cdot \! j_5\, j\, j_5\rangle$ and $\langle \partial\!\cdot \! j_5\, j_5\, j\rangle$ are nonvanishing, but opposite, for repeating the calculation of
\begin{eqnarray}
q^\mu \widetilde T_{\mu\lambda\rho}^{(AAV)}(k_1,k_2) &=& \int
\frac{d^4p }{(2\pi)^{4}}\mathrm{tr}\left\{\frac 1{\slashed{p}}     
\gamma_\lambda\frac 1{\slashed{p}
-\slashed{k}_1}\gamma_{\rho}\gamma_5
\frac 1{\slashed{p}-\slashed{q}}\gamma_5
\slashed{q} \right\}\0\\
& =& \int
\frac{d^4p d^\delta\ell}{(2\pi)^{4+\delta}}\mathrm{tr}\Big\{\frac{\slashed{p}+\slashed{\ell}}{{p}^2-{\ell}^2}     
\gamma_\lambda \frac{\slashed{p}-\slashed{k}_1+\slashed{\ell}}{(p-k_1)^2 -\ell^2} \gamma_{\rho}\gamma_5
\frac{\slashed{p}-\slashed{q}+\slashed{\ell}}{(p-q)^2-\ell^2}\gamma_5\slashed{q} \Big\}
\label{triangleAAV}
\ee
we find 
\be
q^\mu \widetilde T_{\mu\lambda\rho}^{(AAV)}(k_1,k_2)+q^\mu \widetilde T_{\mu\rho\lambda}^{(AAV)}(k_2,k_1)= \frac i{4\pi^2} \Big[\eta_{\lambda\rho}q^2  -\frac 23 (k_{1\lambda} k_{2\rho} -k_{2\lambda} k_{1\rho}\Big]\label{triangleAAV2}
\ee

Thus
\be
\frac 12 \left(\langle \partial\!\cdot \! j_5\, j\, j_5\rangle+ \langle \partial\!\cdot \! j_5\, j_5\, j\rangle
\right)=0 \label{dj5jj5}
\ee

\subsection{Result in coordinate space}

\vskip 1cm
Inserting the above results and, in particular, \eqref{triangleAVA3} into the formula for the effective action
\be
W[V,A]&=& W[0,0]+\sum_{n,m=1}^\infty \frac {i^{n+m-1}}{n!m!} \int \prod_{i=1}^n d^dx_i \,
V_{\mu_i}(x_i) \,\prod_{j=1}^m d^dy_j A_{\nu_j}(y_j)\0\\
&& \times \langle 0|{\cal T} j_{\mu_1}(x_1) \ldots j_{\mu_n}(x_n) j_{5\nu_1}(y_1)\ldots j_{5\nu_m}(x_m) |0\rangle_c \label{dj5jj5} 
\ee
from which one can extract the effective vector current
\be
\langle\!\langle j_\mu(x) \rangle\! \rangle&=&\frac {\delta W[V,A]}{\delta V_\mu(x)}=\sum_{n,m=0}^\infty \frac {i^{n+m}}{n!m!} \int \prod_{i=1}^n d^dx_i \,
V_{\mu_i}(x_i) \,\prod_{j=1}^m d^dy_j A_{\nu_j}(y_j)\0\\
&& \times \langle 0|{\cal T}j_\mu(x) j_{\mu_1}(x_1) \ldots j_{\mu_n}(x_n) j_{5\nu_1}(y_1)\ldots j_{5\nu_m}(x_m) |0\rangle_c \label{JVA}
\ee
and axial current
\be
\langle\!\langle j_{5\mu}(x) \rangle\! \rangle&=&\frac {\delta W[V,A]}{\delta A_\mu(x)}=\sum_{n,m=0}^\infty \frac {i^{n+m}}{n!m!} \int \prod_{i=1}^n d^dx_i \,
V_{\mu_i}(x_i) \,\prod_{j=1}^m d^dy_j A_{\nu_j}(y_j)\0\\
&& \times \langle 0|{\cal T}j_{5\mu}(x) j_{\mu_1}(x_1) \ldots j_{\mu_n}(x_n) j_{5\nu_1}(y_1)\ldots j_{5\nu_m}(x_m) |0\rangle_c \label{J5VA}
\ee
one finds
\be
&&\partial^\mu\langle\!\langle j_\mu(x) \rangle\! \rangle=-\int d^4y d^4z \left(\frac 12 V^\lambda(y) V^\rho(z) \langle0| \partial^\mu j_\mu(x) j_\lambda(y) j_\rho(z)|0\rangle\right. \label{vectordiv}\\
&&+\left.  V^\lambda(y) A^\rho(z) \langle0| \partial^\mu j_\mu(x) j_\lambda(y) j_{5\rho}(z)|0\rangle +\frac 12  A^\lambda(y) A^\rho(z) \langle0| \partial^\mu j_\mu(x) j_{5\lambda}(y) j_{5\rho}(z)|0\rangle\right)=0\0
\ee
and
\be
&&\partial^\mu\langle\!\langle j_{5\mu}(x) \rangle\! \rangle=-\int d^4y d^4z \left(\frac 12 V^\lambda(y) V^\rho(z) \langle0| \partial^\mu j_{5\mu}(x) j_\lambda(y) j_\rho(z)|0\rangle\right. \label{axiladiv}\\
&&+\left.  V^\lambda(y) A^\rho(z) \langle0| \partial^\mu j_{5\mu}(x) j_\lambda(y) j_{5\rho}(z)|0\rangle +\frac 12  A^\lambda(y) A^\rho(z) \langle0| \partial^\mu j_{5\mu}(x) j_{5\lambda}(y) j_{5\rho}(z)|0\rangle\right)=0\0
 \ee
These are the gauge Ward identities to order two in the potentials\footnote{One can easily prove that the WI are satisfied also to order 1 in the potential, i.e. for two-point correlators.}. Of course there are no anomalies. The basic remark here is that there is no ambiguity in passing from the regularization of $\langle0| j_\mu(x) j_\lambda(y) j_\rho(z)|0\rangle$ to the regularization of $\langle0| \partial^\mu j_\mu(x) j_\lambda(y) j_\rho(z)|0\rangle$, in other words $ \partial^\mu\langle0|  j_\mu(x) j_\lambda(y) j_\rho(z)|0\rangle$ is the same as $\langle0| \partial^\mu j_\mu(x) j_\lambda(y) j_\rho(z)|0\rangle$.
As we have seen the situation is different for trace anomalies.

\subsection*{Appendix B. Perturbative cohomology}
\label{ss:pertcoho}

In this Appendix we define the form of local cohomology which is needed in
a perturbative approach. Let us start from the gauge transformations.
\be
\delta A = d\lambda + [A,\lambda], \quad\quad \delta \lambda =-\frac 12
[\lambda,\lambda]_+,\quad\quad \delta^2=0,\quad\quad 
\lambda = \lambda^a(x) T^a \label{gaugetransf}
\ee
To dovetail the perturbative expansion it is useful to split it by considering $A$ and
$\lambda$ infinitesimal (and the latter anticommuting) and define the perturbative cohomology
\be 
&&\delta^{(0)} A = d\lambda, \quad\quad \delta^{(0)}\lambda=0, \quad\quad 
(\delta^{(0)})^2=0\0\\
&&\delta^{(1)} A = [A,\lambda],  \quad\quad  \delta^{(1)}\lambda=-\frac 12
[\lambda,\lambda]_+\0\\
&& \delta^{(0)} \delta^{(1)}+\delta^{(1)} \delta^{(0)}=0,\quad\quad
(\delta^{(1)})^2=0\label{pertcoho}
\ee

The full coboundary operator for diffeomorphisms is given by the transformations
\begin{equation}
\delta_\xi g_{\mu\nu}=\nabla_\mu \xi_\nu + \nabla_\nu \xi_\nu, \quad\quad 
\delta_\xi \xi^\mu = \xi^\lambda \partial_\lambda \xi^\mu\label{fullcoho}
\end{equation}
with $\xi_\mu = g_{\mu\nu}\xi^\nu$.
We can introduce a perturbative cohomology, or graded cohomology, using as
grading the order of infinitesimal, as follows
\begin{eqnarray}
g_{\mu\nu} = \eta_{\mu\nu} + h_{\mu\nu}, \quad\quad g^{\mu\nu}= \eta^{\mu\nu}
-h^{\mu\nu}+ h^\mu_\lambda h^{\lambda\nu}+\ldots \label{metricapprox}
\end{eqnarray}
The analogous expansions for the vielbein is 
\begin{equation}
 e_\mu^a = \delta_\mu^a+\chi_\mu^a+\frac 12 \psi_\mu^a+\ldots,  \0
\end{equation}
Since $e_\mu^a \eta_{ab}e^b_\nu=h_{\mu\nu}$, we can choose
\begin{equation}
\chi_{\mu\nu}= \frac 12 h_{\mu\nu},\quad\quad \psi_{\mu\nu} = 
-\chi_\mu^a\chi_{a\nu}= -\frac 14 h_\mu^\lambda h_{\lambda\nu},\quad\quad \ldots
\label{dreibein}
\end{equation}
  
Inserting the above expansions in (\ref{fullcoho}) we see that we have a grading
in the transformations, given by the order of infinitesimals.
So we can define a  sequence of transformations
\begin{equation}
\delta_\xi= \delta^{(0)}_\xi+\delta^{(1)}_\xi+\delta^{(2)}_\xi+\ldots \nonumber
\end{equation}
At the lowest level we find immediately
\begin{eqnarray}
&& \delta^{(0)}_\xi h_{\mu\nu}= \partial_\mu\xi_\nu + \partial_\nu
\xi_\mu,\quad\quad  \delta^{(0)}_\xi \xi_\mu=0\label{0cohomo}
\end{eqnarray}
and $\xi_\mu = \xi^\mu$. Since $(\delta^{(0)}_\xi)^2=0$ this defines a
cohomology problem. 

At the next level we get
\begin{equation}
\delta_\xi^{(1)} h_{\mu\nu}   =  \xi^\lambda
\partial_\lambda h_{\mu\nu} +
\partial_\mu \xi^\lambda h_{\lambda\nu} + 
\partial_\nu \xi^\lambda h_{\mu\lambda}, \quad \quad \delta^{(1)}_\xi \xi^\mu =
\xi^\lambda \partial_\lambda \xi^\mu\label{1cohomo}
\end{equation}
 
One can verify that
\begin{equation}
 (\delta^{(0)}_\xi)^2=0\quad\quad   \delta^{(0)}_\xi\delta^{(1)}_\xi
+\delta^{(1)}_\xi\delta^{(0)}_\xi =0,\quad\quad
(\delta^{(1)})^2=0\label{xinilpotency}
\end{equation}

Proceeding in the same way we can define an analogous sequence of
transformations for the Weyl transformations. 
From $g_{\mu\nu} = \eta_{\mu\nu} + h_{\mu\nu} $
and $\delta_\omega h_{\mu\nu}= 2\omega g_{\mu\nu}$ we find
\begin{eqnarray}
&&\delta^{(0)}_\omega h_{\mu\nu} = 2\omega \eta_{\mu\nu},\quad\quad
\delta^{(1)}_\omega h_{\mu\nu}= 2\omega h_{\mu\nu},\quad\quad
\delta^{(2)}_\omega h_{\mu\nu}= 0,\ldots \label{deltaomegah}
\end{eqnarray}
as well as $ \delta^{(0)}_\omega \omega=\delta^{(1)}_\omega \omega=0,...$.

Notice that we have 
$\delta^{(0)}_\xi \omega=0, \delta^{(1)}_\xi \omega=\xi^\lambda \partial_\lambda
\omega$.
As a consequence we can extend (\ref{xinilpotency}) to
\begin{equation}
(\delta^{(0)}_\xi +\delta^{(0)}_\omega) (\delta^{(1)}_\xi +\delta^{(1)}_\omega)+
(\delta^{(1)}_\xi +\delta^{(1)}_\omega) (\delta^{(0)}_\xi
+\delta^{(0)}_\omega)=0\label{couplednilpotency}
\end{equation}
and $\delta^{(1)}_\xi\delta^{(1)}_\omega+\delta^{(1)}_\omega
\delta^{(1)}_\xi=0$,
which together with the previous relations make
\begin{equation}
(\delta^{(0)}_\xi +\delta^{(0)}_\omega+\delta^{(1)}_\xi
+\delta^{(1)}_\omega)^2=0 \label{fullcouplednilpotency}
\end{equation}}

%% The bibliography section

\end{document}